\documentstyle[amssymb,thmsa,a4,sw20lart]{article}

\input tcilatex
\begin{document}

\topmargin 0pt \oddsidemargin 5mm

\setcounter{page}{1}

\hspace{8cm}{} \vspace{2cm}

\begin{center}
{\large {EXCITATION OF ELECTROMAGNETIC WAKE FIELDS BY ONE-DIMENSIONAL
ELECTRON BUNCH IN PLASMA IN THE PRESENCE OF CIRCULARLY POLARIZED INTENSE
ELECTROMAGNETIC WAVE}}\\

S.S. Elbakian$^1$ and {H.B. Nersisyan}$^2$\\\vspace{1cm}$^1${\em Yerevan
Physics Institute, Alikhanian Brothers St. 2, Yerevan 375036, Armenia}\\$^2$%
{\em Division of Theoretical Physics, Institute of Radiophysics and
Electronics, Alikhanian Brothers St. 2, Ashtarak-2, 378410, Republic of
Armenia}\footnote{%
E-mail: Hrachya@irphe.am}
\end{center}

\vspace {5mm} \centerline{{\bf{Abstract}}}

The excitation of electromagnetic (EM) wake waves in electron plasma by an
one-dimensional bunch of charged particles has been considered in the
presence of intense monochromatic circularly polarized electromagnetic
(CPEM) pump wave. In the zero state (in the absence of bunch) the
interaction of the pump wave with plasma is described by means of Maxwell
equations and relativistic nonlinear hydrodynamic equations of cold plasma.
The excitation of linear waves by one-dimensional bunch is considered on
this background. It is shown that there are three types of solutions of
linear equations obtained for induced waves corresponding to three ranges of
parameter values of the pump wave, bunch and plasma. In the first range of
parameter values the amplitude of transverse components of induced waves is
shown to grow as the bunch energy and after some value of the relativistic
factor of the bunch to be almost independent of the energy and increase
proportional to the intensity and frequency of the pump wave. The dependence
of longitudinal component of induced waves on the relativistic factor of the
bunch is weak. Its amplitude and wavelength grow as the intensity of pump
wave. The second range of parameters is a resonance one. The amplitude of
the wave excited by the bunch is a linear function of the distance to the
bunch. In the third range of parameter values the longitudinal component of
induced fields are localized near the bunch boundaries and are exponentially
decreased with the increase in distance from these boundaries. The
amplitudes of transverse components of induced waves reach a constant value
with the distance from the bunch boundaries.

{\bf PACS number(s):} 52.35.-g, 52.40.Mj, 52.40.Nk

\newpage 

\section{Introduction}

At present the studies on new methods for charged particle acceleration by
means of wake fields generated in plasma by laser radiation (BWA (Beat Wave
Acceleration), LWFA (Laser Wake Field Acceleration)) and by bunches of
relativistic particles (PWFA (Plasma Wake-Field Acceleration)) in flight
through plasma are intensively developed (see, e.g., the reviews [1, 2] and
references therein). The intensity of acceleration fields (in the order of $%
10^7-10^8V/cm)$, attained by these methods can be used both for the charge
acceleration, and for focusing of electron (positron) bunches in order to
obtain the beams of high density and to ensure high luminosity in linear
colliders of next generation.

The linear theory of wake field generation by rigid bunches of charged
particles in boundless and limited plasma was developed in many works (see,
e.g., [3-11]). The nonlinear theory of wake field generation by a rigid
one-dimensional bunch of finite extent was developed in [12-18]. An
important result of this theory is the proof that the wave breaking limit is 
$E_{\max }=(mv_0\omega _p/e)\left[ 2(\gamma _0-1)\right] ^{1/2}$ and it is
reached at $n_b/n_0\lesssim 1/(2+\gamma _0^{-1})$, where $\omega _p=\left(
4\pi n_0e^2/m\right) ^{1/2}$ is the plasma frequency of electrons, $n_b$ and 
$n_0$ are the densities of the bunch and plasma electrons, $\gamma _0=\left(
1-\beta _0^2\right) ^{-1/2}$ is the relativistic factor of the bunch, $\beta
_0=v_0/c$. The Dawson [17] wave breaking limit is equal to $E_{\max }\approx
2mv_0\omega _p/e$ when $\gamma _0\approx 1$ $(\beta _0\ll 1)$. In the linear
case $n_b/n_0\ll \ll 1$, $E_{\max }\simeq 2(mv_0\omega _p/e)(n_b/n_0)$ for
arbitrary $\gamma _0$.

The one-dimensional relativistic strong waves can be excited in the plasma
by wide relativistic bunches of charged particles or intense laser pulses
[1, 19], (when $k_pa_0\gg 1$, where $k_p=\omega _p/v_0$, $a_0$ are the
characteristic transverse sizes of bunches or pulses).

In the present work the effect of the CPEM pump wave with an arbitrary
intensity parameter ($A=eE_0/mc\omega _0$, where $E_0$ and $\omega _0$ are
the amplitude and the frequency of the EM wave) on the excitation of EM wake
waves by one-dimensional relativistic electron bunch in cold plasma has been
studied. Here the rate of electron oscillations in the pump wave may be of
the order of light velocity. One can obtain the exact solutions of the
Maxwell equations and of nonlinear hydrodynamic equations [18, 20] for CPEM
wave interacting with plasma as well as derive the exact dispersion equation
for waves propagating in the same direction as pump wave [20, 21]. The
parametric instability of the plasma in the presence of CPEM wave has been
studied rather well yet in early works (see, e.g., [20, 22, 23] and
literature therein). Max and Perkins [22] investigated an aperiodic low
frequency instability of plasma in the dipole approximation. The instability
of plasma in the presence of strong CPEM wave was considered in [20] and it
was shown that at parametric excitation of nonpotential oscillations in
plasma by a CPEM the relativistic motion of electrons is essential for the
arbitrary value of pump wave amplitude.

This paper is organized as follows. The derivation of equations for the
velocity of motion of an electron liquid in plasma and for EM fields excited
by an one-dimensional electron bunch in the presence of CPEM wave is given
in Sec. II. The interaction of the pump wave with plasma (in the absence of
a bunch) is described by Maxwell equations and nonlinear hydrodynamic
equations of cold plasma. In this case a spatially homogeneous state of
plasma is possible [18, 20]. Then, assuming that this state is weakly
perturbed by one-dimensional bunch in the plasma, the equations for induced
fields, density and velocity of plasma electrons are obtained using the
perturbation theory methods. In Sec. III the general expressions for induced
EM fields of one-dimensional bunch with an arbitrary density profile are
obtained using the method of Green's functions. There are three ranges of
values of plasma, pump wave and bunch parameters, where the properties of
the Green's functions and, hence, the behavior of excited waves, are
abruptly changed. In Sec. IV the fields excited in plasma by a bunch with
uniform distribution of electron density are considered. For each of the
mentioned ranges of parameters the expressions for induced fields components
are obtained and corresponding numerical calculations are carried out. A
brief summary of the results is given in Sec. V. An alternative method for
obtaining the induced fields is discussed in the Appendix.

\section{Basic Equations}

As an initial system of equations we shall use the Maxwell equations and
relativistic hydrodynamic equations of motion of cold electron plasma under
assumption that the oscillation velocity of plasma electrons in a CPEM wave
much exceeds their thermal velocities and the frequency $\omega _0$ of the
CPEM wave is much higher than the frequency of electron-ion collisions:

\begin{equation}
{\bf \nabla \times B}=\frac 1c\frac{\partial {\bf E}}{\partial t}-\frac{4\pi
e}cn{\bf v-}\frac{4\pi e}c{\bf u}N_b(\xi ),
\end{equation}

\begin{equation}
{\bf \nabla \times E}=-\frac 1c\frac{\partial {\bf B}}{\partial t},\qquad 
{\bf \nabla B}=0,
\end{equation}

\begin{equation}
{\bf \nabla E}=-4\pi e\left( n-n_0\right) -4\pi eN_b(\xi ),
\end{equation}

\begin{equation}
\frac{\partial {\bf v}}{\partial t}+\left( {\bf v\nabla }\right) {\bf v}%
=-\frac em\sqrt{1-\frac{v^2}{c^2}}\left[ {\bf E}+\frac 1c{\bf v}\times {\bf B%
}-\frac{{\bf v}}{c^2}\left( {\bf vE}\right) \right] ,
\end{equation}

\begin{equation}
\frac{\partial n}{\partial t}+{\bf \nabla }\left( n{\bf v}\right) =0,
\end{equation}
where $n_0$ is the unperturbed electron density, $N_b(\xi )$ is the density
of one-dimensional electron bunch moving with velocity ${\bf u}=u{\bf e}_z$, 
$\xi =z-ut$.

In the CPEM wave propagating along the $z$ axis a spatially homogeneous
state of plasma may be established, in which the EM field and velocity of
electrons will be determined by expressions [18, 20]

\begin{equation}
{\bf E}_0=E_0\left( {\bf e}_x\cos \zeta +{\bf e}_y\sin \zeta \right) ,\qquad
E_{0z}=0,
\end{equation}

\begin{equation}
{\bf B}_0=\frac{k_0c}{\omega _0}E_0\left( -{\bf e}_x\sin \zeta +{\bf e}%
_y\cos \zeta \right) ,\qquad B_{0z}=0,
\end{equation}

\begin{equation}
{\bf v}_e=c\beta _e\left( -{\bf e}_x\sin \zeta +{\bf e}_y\cos \zeta \right)
,\qquad v_{ez}=0,
\end{equation}
where $\zeta =\omega _0t-k_0z$, $k_0=(\omega _0/c)\sqrt{\varepsilon (\omega
_0)}$, $\varepsilon (\omega )=1-\omega _L^2/\omega ^2$, $\omega _L^2=\omega
_p^2\sqrt{1-\beta _e^2}$, $\beta _e=v_e/c$, $A=eE_0/mc\omega _0$,

\begin{equation}
v_e=c\frac A{\sqrt{1+A^2}},
\end{equation}
$\omega _p^2=4\pi n_0e^2/m$ is the plasma frequency, $c$ is the velocity of
light.

Consider small perturbations of plasma due to the presence of an electron
bunch with density of $N_b$. The linearization implies that the analysis is
valid only on condition that the bunch density $N_b\ll n_0$. We shall write
all quantities in the form $f=f_0+f^{^{\prime }}$, where $f_0$ are
determined by means of Eqs. (6)-(9). The linearization of the system of Eqs.
(1)-(5) gives the following set of equations for induced variables $%
f^{^{\prime }}$

\begin{equation}
{\bf \nabla \times B}^{^{\prime }}=\frac 1c\frac{\partial {\bf E}^{^{\prime
}}}{\partial t}-\frac{4\pi e}c\left( n_0{\bf v}^{^{\prime }}+n^{^{\prime }}%
{\bf v}_e\right) {\bf -}\frac{4\pi e}c{\bf u}N_b(\xi ),
\end{equation}

\begin{equation}
{\bf \nabla \times E}^{^{\prime }}=-\frac 1c\frac{\partial {\bf B}^{^{\prime
}}}{\partial t},\qquad {\bf \nabla B}^{^{\prime }}=0,
\end{equation}

\begin{equation}
{\bf \nabla E}^{^{\prime }}=-4\pi en^{^{\prime }}-4\pi eN_b(\xi ),
\end{equation}

\begin{eqnarray}
&&\frac{\partial {\bf v}^{^{\prime }}}{\partial t}+\left( {\bf v}_e{\bf %
\nabla }\right) {\bf v}^{^{\prime }}+\left( {\bf v}^{^{\prime }}{\bf \nabla }%
\right) {\bf v}_e \\
&=&-\frac em\sqrt{1-\beta _e^2}\left[ {\bf E}^{^{\prime }}+\frac 1c{\bf v}%
_e\times {\bf B}^{^{\prime }}+\frac 1c{\bf v}^{^{\prime }}\times {\bf B}_0-%
\frac{{\bf v}_e}{c^2}\left( {\bf v}_e{\bf E}^{^{\prime }}+{\bf v}^{^{\prime
}}{\bf E}_0\right) \right] +  \nonumber \\
&&+\frac{e\left( {\bf v}^{^{\prime }}{\bf v}_e\right) }{mc^2\sqrt{1-\beta
_e^2}}\left( {\bf E}_0+\frac 1c{\bf v}_e\times {\bf B}_0\right) ,  \nonumber
\end{eqnarray}

\begin{equation}
\frac{\partial n^{^{\prime }}}{\partial t}+n_0{\bf \nabla v}^{^{\prime
}}+\left( {\bf v}_e{\bf \nabla }\right) n^{^{\prime }}=0.
\end{equation}
The set of Eqs. (10)-(14) is a system of partial differential equations with
periodical coefficients with respect to the variable $\zeta $.

If we pass from $x$ and $y$ components of fields and velocities of plasma
electrons to new variables

\begin{equation}
\left( 
\begin{tabular}{l}
$E^{\pm }$ \\ 
$B^{\pm }$ \\ 
$w^{\pm }$%
\end{tabular}
\right) =\left( 
\begin{tabular}{l}
$E_x^{^{\prime }}\pm iE_y^{^{\prime }}$ \\ 
$B_x^{^{\prime }}\pm iB_y^{^{\prime }}$ \\ 
$v_x^{^{\prime }}\pm iv_y^{^{\prime }}$%
\end{tabular}
\right) =\left( 
\begin{tabular}{l}
${\cal E}^{\pm }$ \\ 
${\cal B}^{\pm }$ \\ 
${\cal V}^{\pm }$%
\end{tabular}
\right) e^{i\zeta }
\end{equation}
this will mean that we convert these into the rotating system of reference
of the CPEM wave. Here the set of Eqs. (10)-(14) passes into a system of
ordinary inhomogeneous differential equations with constant coefficients

\begin{eqnarray}
&&\frac 1{\gamma ^2}\frac{\partial ^2{\cal E}^{\pm }}{\partial \xi ^2}\mp
2i\left( k_0-\beta \frac{\omega _0}c\right) \frac{\partial {\cal E}^{\pm }}{%
\partial \xi }-\left( k_0^2-\frac{\omega _0^2}{c^2}\right) {\cal E}^{\pm } \\
&=&\frac{4\pi en_0}c\left( \beta \frac{\partial {\cal V}^{\pm }}{\partial
\xi }\mp i\frac{\omega _0}c{\cal V}^{\pm }+\omega _0\beta _e\frac{%
n^{^{\prime }}}{n_0}\pm \frac{i\beta _e}{n_0}u\frac{\partial n^{^{\prime }}}{%
\partial \xi }\right) ,  \nonumber
\end{eqnarray}

\begin{eqnarray}
&&\frac 1{\gamma ^2}\frac{\partial ^2{\cal B}^{\pm }}{\partial \xi ^2}\mp
2i\left( k_0-\beta \frac{\omega _0}c\right) \frac{\partial {\cal B}^{\pm }}{%
\partial \xi }-\left( k_0^2-\frac{\omega _0^2}{c^2}\right) {\cal B}^{\pm } \\
&=&\pm \frac{4\pi ein_0}c\left( \frac{\partial {\cal V}^{\pm }}{\partial \xi 
}\mp ik_0{\cal V}^{\pm }+k_0v_e\frac{n^{^{\prime }}}{n_0}\pm \frac{iv_e}{n_0}%
\frac{\partial n^{^{\prime }}}{\partial \xi }\right) ,  \nonumber
\end{eqnarray}

\begin{eqnarray}
\frac{\partial ^2{\cal V}^{\pm }}{\partial \xi ^2}+\frac{\omega _0^2}{u^2}%
{\cal V}^{\pm } &=&\frac e{mu}\sqrt{1-\beta _e^2} \\
&&\left[ \left( 1-\frac{\beta _e^2}2\right) \left( \frac{\partial {\cal E}%
^{\pm }}{\partial \xi }\pm i\frac{\omega _0}u{\cal E}^{\pm }\right) +\frac{%
\beta _e^2}2\left( \frac{\partial {\cal E}^{\mp }}{\partial \xi }\mp i\frac{%
\omega _0}u{\cal E}^{\mp }\right) \right] ,  \nonumber
\end{eqnarray}

\begin{equation}
\frac{\partial E_z^{^{\prime }}}{\partial \xi }=-4\pi en^{^{\prime }}-4\pi
eN_b(\xi ),\qquad B_z^{\prime }=0,
\end{equation}

\begin{equation}
\frac{\partial v_z^{^{\prime }}}{\partial \xi }=\frac e{mu}\sqrt{1-\beta _e^2%
}\left[ E_z^{^{\prime }}-\frac{\beta _e}2\left( {\cal B}^{+}+{\cal B}%
^{-}\right) +\frac{k_0E_0}{2\omega _0}\left( {\cal V}^{+}+{\cal V}%
^{-}\right) \right] ,
\end{equation}

\begin{equation}
n^{^{\prime }}=n_0\frac{v_z^{^{\prime }}}u,
\end{equation}
where $\beta =u/c.$ At the derivation of Eqs. (16)-(21) we assumed that all
quantities depended on the variable $\xi =z-ut$. One can see from these
equations that owing to the presence of CPEM wave in the plasma, the induced
fields and velocities of electron liquid motion are parametrically related.
Besides the longitudinal component of the induced electric field the
right-hand side of Eq. (20) for the induced velocity longitudinal component
comprises also the Lorentz force. The second term arises on account of the
interaction of unperturbed velocity of plasma electrons with the induced
magnetic field. The third term is due to the interaction of induced
velocities of electrons with the magnetic field of CPEM wave. As is seen
from the expression (20), the coefficient of the second term ($\beta _e$) in
the square brackets tends to $1/2\,\,$and the coefficient of the last term
grows with the intensity of the CPEM wave. Below we shall examine the
solutions of Eqs. (16)-(21).

\section{Calculations of Green's Functions}

To solve the system of Eqs. (16)-(21) one can obtain the equations for each
of variables $n^{^{\prime }}$, $\,E_z^{^{\prime }}$, ${\cal V}^{\pm }$, $%
{\cal B}^{\pm }$, ${\cal E}^{\pm }$. Such an approach to determination of
induced fields and velocities of electrons has its advantages and is
discussed in short in the Appendix. Here we shall solve the system of Eqs.
(16)-(21) by expansion of induced quantities in Fourier integral in variable 
$\xi .$ Then, after some transformations we find

\begin{equation}
\left( 
\begin{tabular}{l}
$E_z^{^{\prime }}(\xi )$ \\ 
${\cal E}^{+}(\xi )$ \\ 
${\cal B}^{+}(\xi )$%
\end{tabular}
\right) =\int_{-\infty }^{+\infty }d\xi ^{^{\prime }}N_b(\xi ^{^{\prime
}})\left( 
\begin{tabular}{l}
$G_z^{(e)}(\xi ^{^{\prime }}-\xi )$ \\ 
$G_{\bot }^{(e)}(\xi ^{^{\prime }}-\xi )$ \\ 
$G_{\bot }^{(m)}(\xi ^{^{\prime }}-\xi )$%
\end{tabular}
\right) ,
\end{equation}
where $G_z^{(e)}$,\thinspace $G_{\bot }^{(e)}$, $G_{\bot }^{(m)}$ are the
Green's functions respectively for quantities $E_z^{^{\prime }}$, ${\cal E}%
^{+}$ and ${\cal B}^{+}$ that are as follows:

\begin{equation}
G_z^{(e)}(s)=-2ie\int_{-\infty }^{+\infty }\frac{dk}k\frac{D_1(k,\omega )}{%
D(k,\omega )}e^{iks},
\end{equation}

\begin{equation}
G_{\bot }^{(e)}(s)=-\frac{2e\beta _e\omega _L^2}c\int_{-\infty }^{+\infty
}dk(ku+\omega _0)\frac{R_{-1}(k,\omega )}{D(k,\omega )}e^{iks},
\end{equation}

\begin{equation}
G_{\bot }^{(m)}(s)=-2ie\beta _e\omega _L^2\int_{-\infty }^{+\infty }dk(k+k_0)%
\frac{R_{-1}(k,\omega )}{D(k,\omega )}e^{iks},
\end{equation}
where $\omega =ku$, $\gamma ^{-2}=1-\beta ^2$. Here we made the following
notations:

\begin{equation}
R_{\pm 1}(k,\omega )=\left( k\pm k_0\right) ^2-\frac{\left( \omega \pm
\omega _0\right) ^2}{c^2}\varepsilon (\omega \pm \omega _0),
\end{equation}

\begin{equation}
D_1(k,\omega )=\omega ^2R_1(k,\omega )R_{-1}(k,\omega )+\frac{\beta
_e^2\omega _L^2}2\left( k^2-\frac{\omega ^2}{c^2}\right) \left[ R_1(k,\omega
)+R_{-1}(k,\omega )\right] ,
\end{equation}

\begin{equation}
D(k,\omega )=\omega ^2\varepsilon (\omega )R_1(k,\omega )R_{-1}(k,\omega )+%
\frac{\beta _e^2\omega _L^2}2\left[ k^2-\frac{\omega ^2}{c^2}\varepsilon
(\omega )\right] \left[ R_1(k,\omega )+R_{-1}(k,\omega )\right] .
\end{equation}
From expressions (15) one can find the transverse components of induced
electric and magnetic fields, as well as the velocities of plasma electron
motion by taking the real or imaginary parts from complex quantities ${\cal E%
}^{+}$, ${\cal B}^{+}$ and $w^{+}$. As a result we have

\begin{equation}
\left( 
\begin{tabular}{l}
$E_x^{^{\prime }}(z,t)$ \\ 
$B_x^{^{\prime }}(z,t)$%
\end{tabular}
\right) =\left( 
\begin{tabular}{l}
$E_r(\xi )$ \\ 
$B_r(\xi )$%
\end{tabular}
\right) \cos (\zeta )-\left( 
\begin{tabular}{l}
$E_i(\xi )$ \\ 
$B_i(\xi )$%
\end{tabular}
\right) \sin (\zeta ),
\end{equation}

\begin{equation}
\left( 
\begin{tabular}{l}
$E_y^{^{\prime }}(z,t)$ \\ 
$B_y^{^{\prime }}(z,t)$%
\end{tabular}
\right) =\left( 
\begin{tabular}{l}
$E_r(\xi )$ \\ 
$B_r(\xi )$%
\end{tabular}
\right) \sin (\zeta )+\left( 
\begin{tabular}{l}
$E_i(\xi )$ \\ 
$B_i(\xi )$%
\end{tabular}
\right) \cos (\zeta ),
\end{equation}
where

\begin{equation}
\left( 
\begin{tabular}{l}
$E_r(\xi )$ \\ 
$B_r(\xi )$%
\end{tabular}
\right) ={\rm Re}\left( 
\begin{tabular}{l}
${\cal E}^{+}(\xi )$ \\ 
${\cal B}^{+}(\xi )$%
\end{tabular}
\right) ,\qquad \left( 
\begin{tabular}{l}
$E_i(\xi )$ \\ 
$B_i(\xi )$%
\end{tabular}
\right) ={\rm Im}\left( 
\begin{tabular}{l}
${\cal E}^{+}(\xi )$ \\ 
${\cal B}^{+}(\xi )$%
\end{tabular}
\right) .
\end{equation}

Thus, one can see from expressions (29) and (30) that the transverse
components of induced fields describe the modulated oscillations in plasma.
E.g., one can represent $E_y^{^{\prime }}$ in the form

\begin{equation}
E_y^{^{\prime }}(z,t)=E_{\perp 0}(\xi )\sin \left( \zeta +\psi _0(\xi
)\right) .
\end{equation}
Here $E_{\bot 0}(\xi )=\sqrt{E_r^2(\xi )+E_i^2(\xi )}$ is the amplitude and $%
\psi _0(\xi )=\arctan [E_i(\xi )/E_r(\xi )]$ is the phase displacement of
oscillations (other induced quantities may be similarly written in the form
of (32)). In the system of reference of bunch rest the expression (32)
describes the transverse harmonic wave. In an arbitrary reference system the
expression (32) describes a modulated transverse wave, the profile of which
is given by the function $E_{\bot 0}(\xi )$. Note also that in the absence
of CPEM wave ($\beta _e=0$) all transverse components of induced quantities
turn to zero, and Eq. (19) and its solution (the first expression in Eq.
(22) together with expressions (27) and (28)) turn into the known
expressions for linear one-dimensional fields.

Now calculate the Green's functions that are determined by expressions
(23)-(25). The poles of integrals in the expressions (23)-(25) are
determined by the dispersion equation $D(k,\omega )=0$, that has been
comprehensively studied for an arbitrary case in Ref. [20] (i.e., without
the $\stackrel{\vee }{\text{C}}$herenkov condition $\omega =ku$ to be
imposed). In the absence of a CPEM wave ($\beta _e=0$) we obtain dispersion
equations for ordinary plasma and transverse (EM) waves from the expression
(28) with $\omega =\omega _p$ and $\omega ^2=\omega _p^2+k^2c^2$\thinspace
respectively. In the presence of a pump wave ($\beta _e\neq 0$) there arise
coupling waves in the plasma, the growth increment of which for small
amplitudes of the CPEM wave ($\beta _e\ll 1$) linearly increases as $\beta
_e $. Thus, the parametric instability of coupling waves persists down to
the value of $\beta _e=0$. However, in the case of non-dense plasma ($\omega
_0\lesssim 10^{13}\sec ^{-1}$ or $n_0\lesssim 10^{17}$cm$^{-3})$ and for
optical values of $\omega _0$ ($\omega _0\sim 10^{15}\sec ^{-1}$), the time
of relativistic bunch interaction is much less than the time of parametric
instability development [20] and here we shall not make any allowance for
their effect on the excitation of waves in plasma.

In case when the $\stackrel{\vee }{\text{C}}$herenkov condition is
approached ($\omega =ku)$ we obtain from expression (28) the following
dispersion equation

\begin{equation}
\left( k^2-\frac{\omega _L^2}{u^2}\right) \left[ k^2-4\gamma ^4\left(
k_0-\beta \frac{\omega _0}c\right) ^2\right] +\frac{\beta _e^2\omega _L^2}{%
u^2}\left( k^2+\gamma ^2\frac{\omega _L^2}{c^2}\right) =0.
\end{equation}
If we introduce a dimensionless wave vector $\lambda $ that is related to $k$
by means of expression $k=(\omega _L/u)\lambda $, then the solution of Eq.
(3) will take on the form

\begin{equation}
\lambda _{\pm }^2=\frac 1{2a^4}+2\beta ^2\gamma ^4F^2\pm \sqrt{\left( \frac
1{2a^4}+2\beta ^2\gamma ^4F^2\right) ^2-\beta ^2\gamma ^2\left( \frac{a^4-1}{%
a^4}+4\gamma ^2F^2\right) },
\end{equation}
where

\begin{equation}
F=\sqrt{a^2\Delta ^2-1}-\beta a\Delta ,
\end{equation}
$\Delta =\omega _0/\omega _p$, $a^2=\sqrt{1+A^2}$.

The behavior of solutions of Eq. (33) and, consequently, the nature of
induced fields and of velocity of plasma electrons strongly depend on the
sign of radicand in the Eq. (34). In the following we shall identify the
regions where the radicand is positive, equal to zero or negative as the I$^{%
\text{st}}$, II$^{\text{nd}}$, and III$^{\text{rd}}$ regions respectively.
First, we shall determine the boundaries of the region where this radicand
turns to zero. If the radicand is made equal to zero one finds for the
boundary of the above-mentioned region

\begin{equation}
\Delta =\frac \gamma a\left( \sqrt{1+\gamma ^2F_{+}^2}\pm \beta \gamma
F_{+}\right) ,
\end{equation}
when $a>1$, and

\begin{equation}
\Delta =\frac \gamma a\left( \sqrt{1+\gamma ^2F_{-}^2}\pm \beta \gamma
F_{-}\right) ,
\end{equation}
when $1<a<a_0(\gamma )$, where

\begin{equation}
F_{\pm }=\sqrt{\frac{2-1/a^4\pm 2\gamma \sqrt{1-1/a^4}}{4\gamma ^2\left(
\gamma ^2-1\right) }},
\end{equation}

\begin{equation}
a_0(\gamma )=\left\{ \frac{\gamma +\sqrt{\gamma ^2-1}}{2\sqrt{\gamma ^2-1}}%
\right\} ^{1/4}.
\end{equation}
The Eqs. (36)-(39) were obtained on the assumption that $\gamma >\gamma
_1\cong 1.45$, where $\gamma _1$ is a real positive root of equation $%
2\gamma ^2(\gamma ^2-2)=\gamma -1$ that satisfies the condition $\gamma >1$.

For small pump wave amplitude ($a-1\ll 1$) the values of functions given by
the first expressions (with plus sign) and those of functions given by the
second expressions (with minus sign) of Eqs. (36) and (37) coincide. We
denote these values respectively as $\Delta _{\pm }$ and from Eqs. (36) and
(37) we can obtain for them the following expressions:

\begin{equation}
\Delta _{\pm }=\gamma \sqrt{1+\frac 1{4(\gamma ^2-1)}}\pm \frac 12.
\end{equation}

In Figs. 1, 2, and 3 the regions I, II and III (the lines coincide with the
region II) are given for values of $\gamma =1.5$, $\gamma =20$ and $\gamma
=100$ respectively. The lines in these figures are closed at the infinity
(i.e., when $a\rightarrow \infty )$. For modern lasers with intensities $%
I_L\lesssim 10^{20}$W$/$cm$^2$ we have the following restriction on the
parameter $a$: $a\lesssim 3$ for values of frequency of about $\omega
_0\simeq 3\times 10^{15}$sec$^{-1}$. So, as is seen from Figs. 1, 2 and 3
and Eqs. (36) and (37), for values of parameters $\omega _0\simeq 10^{15}$sec%
$^{-1}$, $n_0\lesssim 10^{17}$cm$^{-3}$ ($\omega _p\lesssim 10^{13}$sec$%
^{-1} $) the solutions of Eq. (35) are in the region I for wide spread of
the values of $\gamma $ (up to the values of $\gamma \lesssim (\omega
_0/\omega _p)a$ and higher). The solutions of Eq. (33) will be in the
regions III or II if the condition $\gamma \sim a\Delta =(\omega _0/\omega
_p)a\gtrsim 100$ is observed (the ultrarelativistic bunch).

Now we shall calculate the Green's function for each of the regions in
separate.

\subsection{Region I}

In this case the roots of Eq. (33) are real and lie within the upper half
plane of the complex variable $k$. Integrating the expressions (23)-(25)
over the variable $k$ we find

\begin{eqnarray}
G_z^{(e)}(s) &=&\frac{4\pi e\theta (s)}{\lambda _{+}^2-\lambda _{-}^2}\left[
\left( \lambda _{+}^2+\frac{a^4-1}{a^4}-4\beta ^2\gamma ^4F^2\right) \cos
\left( \frac{\lambda _{+}}ak_ps\right) -\right. \\
&&\left. -\left( \lambda _{-}^2+\frac{a^4-1}{a^4}-4\beta ^2\gamma
^4F^2\right) \cos \left( \frac{\lambda _{-}}ak_ps\right) \right] ,  \nonumber
\end{eqnarray}

\begin{eqnarray}
G_{\bot }^{(e)}(s) &=&\pi ie\frac{\sqrt{a^4-1}}a\frac{\Delta F}{F^2+\frac{%
a^4-1}{4a^4}\gamma ^{-2}}\left[ \theta (s)-\theta (-s)\right] + \\
&&+\frac{4\pi e\beta \gamma ^2}{\lambda _{+}^2-\lambda _{-}^2}\frac{\sqrt{%
a^4-1}}{a^2}\theta (s)\left\{ \left( a\Delta -2\beta \gamma ^2F\right)
\left[ \frac 1{\lambda _{+}}\sin \left( \frac{\lambda _{+}}ak_ps\right)
-\frac 1{\lambda _{-}}\sin \left( \frac{\lambda _{-}}ak_ps\right) \right]
+\right.   \nonumber \\
&&\left. +i\left( 1-\frac{2\beta \gamma ^2a\Delta F}{\lambda _{-}^2}\right)
\cos \left( \frac{\lambda _{-}}ak_ps\right) -i\left( 1-\frac{2\beta \gamma
^2a\Delta F}{\lambda _{+}^2}\right) \cos \left( \frac{\lambda _{+}}%
ak_ps\right) \right\} ,  \nonumber
\end{eqnarray}

\begin{eqnarray}
G_{\bot }^{(m)}(s) &=&-\pi e\frac{\sqrt{a^4-1}}{a^2}\frac{FF_0}{F^2+\frac{%
a^4-1}{4a^4}\gamma ^{-2}}\left[ \theta (s)-\theta (-s)\right] +\frac{4\pi
e\gamma ^2}{\lambda _{+}^2-\lambda _{-}^2}\frac{\sqrt{a^4-1}}{a^2}\theta
(s)\times  \\
&&\times \left\{ \left( 1-\frac{2\beta ^2\gamma ^2FF_0}{\lambda _{+}^2}%
\right) \cos \left( \frac{\lambda _{+}}ak_ps\right) -\left( 1-\frac{2\beta
^2\gamma ^2FF_0}{\lambda _{-}^2}\right) \cos \left( \frac{\lambda _{-}}%
ak_ps\right) -\right.   \nonumber \\
&&\left. -i\beta \left( 2\gamma ^2F-F_0\right) \left[ \frac 1{\lambda
_{+}}\sin \left( \frac{\lambda _{+}}ak_ps\right) -\frac 1{\lambda _{-}}\sin
\left( \frac{\lambda _{-}}ak_ps\right) \right] \right\} ,  \nonumber
\end{eqnarray}
where $k_p=\omega _p/u$, $F_0=\sqrt{a^2\Delta ^2-1}$, $\theta (s)$ is the
Heavyside function. It is seen from Eqs. (41)-(43) that the bunch excites in
plasma the oscillations of two types with frequencies $\omega _L\lambda _{+}$
and $\omega _L\lambda _{-}$. In the $a=1$ limit (the CPEM wave is absent) $%
G_{\bot }^{(e)}$ and $G_{\bot }^{(m)}$ turn to zero and for $G_z^{(e)}$ we
obtain the expression

\begin{equation}
G_z^{(e)}(s)=4\pi e\theta (s)\cos (k_ps),
\end{equation}
that coincides with those given in Refs. [24, 25].

\subsection{Region II}

In the second region the radicand in Eq. (34) turns zero. In this case $%
\lambda _{+}^2=\lambda _{-}^2=\lambda _0^2$,

\begin{equation}
\lambda _0^2=\frac 1{2a^4}+2\beta ^2\gamma ^4F^2=1\pm \gamma \frac{\sqrt{%
a^4-1}}{a^2}>0,
\end{equation}
where the sign ''+'' corresponds to the values of parameters $1<a$, $\gamma
>\gamma _1$ (see the Eq. (36)). The sign ''--'' corresponds to the values of
parameters $1<a<a_0(\gamma )$, $\gamma >\gamma _1$ (see the Eq. (37)). The
poles $\pm \lambda _0$ in this region are multiple ones and lie in the upper
complex plane $k$. The calculation of integrals in Eqs. (23)-(25) gives

\begin{equation}
G_z^{(e)}(s)=4\pi e\theta (s)\left[ \cos \left( \frac{\lambda _0}%
ak_ps\right) -\frac{k_ps}{2a\lambda _0}\left( \lambda _0^2+\frac{a^4-1}{a^4}%
-4\beta ^2\gamma ^4F^2\right) \sin \left( \frac{\lambda _0}ak_ps\right)
\right] ,
\end{equation}

\begin{eqnarray}
G_{\bot }^{(e)}(s) &=&\pi ie\frac{\sqrt{a^4-1}}a\frac{\Delta F}{F^2+\frac{%
a^4-1}{4a^4}\gamma ^{-2}}\left[ \theta (s)-\theta (-s)\right] - \\
&&-\frac{2\pi e\beta \gamma ^2}{\lambda _0^3}\frac{\sqrt{a^4-1}}{a^2}\theta
(s)\left\{ \left[ a\Delta -2\beta \gamma ^2F+i\left( 2\beta \gamma ^2\Delta
F-\frac{\lambda _0^2}a\right) (k_ps)\right] \sin \left( \frac{\lambda _0}%
ak_ps\right) -\right.   \nonumber \\
&&\left. -\left[ \left( a\Delta -2\beta \gamma ^2F\right) \frac{\lambda _0}%
a(k_ps)-\frac{4i\beta \gamma ^2a\Delta F}{\lambda _0}\right] \cos \left( 
\frac{\lambda _0}ak_ps\right) \right\} ,  \nonumber
\end{eqnarray}

\begin{eqnarray}
G_{\bot }^{(m)}(s) &=&-\pi e\frac{\sqrt{a^4-1}}{a^2}\frac{FF_0}{F^2+\frac{%
a^4-1}{4a^4}\gamma ^{-2}}\left[ \theta (s)-\theta (-s)\right] +\frac{2\pi
e\gamma ^2}{\lambda _0^3}\frac{\sqrt{a^4-1}}{a^2}\theta (s)\times \\
&&\times \left\{ \left[ \frac{4\beta ^2\gamma ^2FF_0}{\lambda _0}-i\beta
\left( 2\gamma ^2F-F_0\right) \frac{\lambda _0}a(k_ps)\right] \cos \left( 
\frac{\lambda _0}ak_ps\right) +\right.  \nonumber \\
&&\left. +\left[ \frac{k_ps}a\left( 2\beta ^2\gamma ^2FF_0-\lambda
_0^2\right) +i\beta \left( 2\gamma ^2F-F_0\right) \right] \sin \left( \frac{%
\lambda _0}ak_ps\right) \right\} .  \nonumber
\end{eqnarray}
Note, that one could obtain the Eqs. (46)-(48) from expressions (41)-(43) by
tending $\lambda _{+}\rightarrow \lambda _{-}\rightarrow \lambda _0$ in the
latter. The uncertainty that arises here may be eliminated by using the
L'Hopital rule.

Unlike the Eqs. (41)-(43) the Eqs. (46)-(48) contain oscillating terms, the
amplitudes of which are linear functions of $s$. We shall see below that
such a dependence of Green's functions leads to an exponential growth of the
induced fields as a function of coordinates.

\subsection{Region III}

In this case the roots of dispersion Eq. (33) are complex quantities and are
distributed over the complex plane $\lambda $ symmetrical with respect to
the origin of coordinates. We shall denote the roots in the upper half plane
as $\lambda _{\pm }=\pm \alpha +i\delta $ (the roots in the lower complex
half plane will be $\pm \alpha -i\delta $), where

\begin{equation}
\left( 
\begin{tabular}{l}
$\alpha $ \\ 
$\beta $%
\end{tabular}
\right) =\sqrt{\frac{\sqrt{P_1^2+P_2^2}\pm P_1}2},
\end{equation}

\begin{equation}
P_1=\frac 1{2a^4}+2\beta ^2\gamma ^4F^2,
\end{equation}

\begin{equation}
P_2=\sqrt{\beta ^2\gamma ^2\left( \frac{a^4-1}{a^4}+4\gamma ^2F^2\right)
-P_1^2}.
\end{equation}

After calculation of the integrals over $k$ in Eqs. (23)-(25) we have the
following expressions for the Green's functions:

\begin{eqnarray}
G_z^{(e)}(s) &=&\pi e\left[ \theta (s)-\theta (-s)\right] \exp \left( -\frac
\delta ak_p|s|\right)  \\
&&\times \left[ \frac 1{\alpha \delta }\left( \alpha ^2-\delta ^2+\frac{a^4-1%
}{a^4}-4\beta ^2\gamma ^4F^2\right) \sin \left( \frac \alpha ak_p|s|\right)
+2\cos \left( \frac \alpha ak_ps\right) \right] ,  \nonumber
\end{eqnarray}

\begin{eqnarray}
G_{\bot }^{(e)}(s) &=&\pi ie\frac{\sqrt{a^4-1}}a\frac{\Delta F}{F^2+\frac{%
a^4-1}{4a^4}\gamma ^{-2}}\left[ \theta (s)-\theta (-s)\right] -\frac{\pi
e\beta \gamma ^2}{\alpha \delta \left( \alpha ^2+\delta ^2\right) }\frac{%
\sqrt{a^4-1}}{a^2}\exp \left( -\frac \delta ak_p|s|\right) \times \\
&&\times \left\{ \alpha \left[ \left( a\Delta -2\beta \gamma ^2F\right) 
\frac{|s|}s+\frac{4i\delta \beta \gamma ^2a\Delta F}{\alpha ^2+\delta ^2}%
\right] \cos \left( \frac \alpha ak_ps\right) +\right.  \nonumber \\
&&\left. +\left[ \delta \left( a\Delta -2\beta \gamma ^2F\right) \frac{|s|}%
s+i\frac{\left( \alpha ^2+\delta ^2\right) ^2-2\beta \gamma ^2a\Delta
F\left( \alpha ^2-\delta ^2\right) }{\alpha ^2+\delta ^2}\right] \sin \left(
\frac \alpha ak_p|s|\right) \right\} ,  \nonumber
\end{eqnarray}

\begin{eqnarray}
G_{\bot }^{(m)}(s) &=&-\pi e\frac{\sqrt{a^4-1}}{a^2}\frac{FF_0}{F^2+\frac{%
a^4-1}{4a^4}\gamma ^{-2}}\left[ \theta (s)-\theta (-s)\right] +\frac{\pi
e\gamma ^2}{\alpha \delta \left( \alpha ^2+\delta ^2\right) }\frac{\sqrt{%
a^4-1}}{a^2}\exp \left( -\frac \delta ak_p|s|\right) \times \\
&&\times \left\{ \alpha \beta \left[ \frac{4\delta \beta \gamma ^2FF_0}{%
\alpha ^2+\delta ^2}\frac{|s|}s+i\left( 2\gamma ^2F-F_0\right) \right] \cos
\left( \frac \alpha ak_ps\right) +\right.  \nonumber \\
&&\left. +\left[ \frac{\left( \alpha ^2+\delta ^2\right) ^2-2\beta ^2\gamma
^2FF_0\left( \alpha ^2-\delta ^2\right) }{\alpha ^2+\delta ^2}\frac{|s|}%
s+i\beta \delta \left( 2\gamma ^2F-F_0\right) \right] \sin \left( \frac
\alpha ak_p|s|\right) \right\} .  \nonumber
\end{eqnarray}

It follows from Eqs. (52)-(54) that in the region III the Green's functions
contain terms that oscillate with exponentially decreasing amplitude.

\section{Induced Fields for Specific Choice of $N_b(\xi )$}

In this section we shall calculate and investigate the induced fields for a
specific profile of electron bunch density. We shall assume that the
electrons are homogeneously distributed in the bunch with the density $n_b$ $%
(n_{b\ll }n_0)$, and the length of bunch is $d$, i.e.,

\begin{equation}
N_b(\xi )=n_b\left[ \theta (\xi )-\theta (\xi -d)\right] .
\end{equation}

As was noted in Sec. III, the values of induced fields strongly depend on
the fact, in which of three mentioned regions lie the values of the
discriminant of Eq. (33). The induced fields in each region (I, II or III)
of the values of bunch, plasma and pump wave parameters will be considered
separately.

\subsection{Region I}

In this case the Green's functions are determined by the Eqs. (41)-(43). The
substitution of these expressions and Eq. (55) into the Eq. (22) will result
in the obtaining of following expressions for the induced fields ahead of ($%
\xi >d$), inside ($0<\xi <d$) and behind ($\xi <0$) the bunch:

$\xi >d;$

\begin{equation}
E_z^{^{\prime }}(\xi )=0,
\end{equation}

\begin{equation}
{\cal E}^{+}(\xi )=-i\widetilde{E}_0\frac{n_b}{n_0}(k_pd)\frac{\sqrt{a^4-1}}a%
\frac{\beta \Delta F}{4F^2+\frac{a^4-1}{a^4}\gamma ^{-2}},
\end{equation}

\begin{equation}
{\cal B}^{+}(\xi )=\widetilde{E}_0\frac{n_b}{n_0}\left( k_pd\right) \frac{%
\sqrt{a^4-1}}{a^2}\frac{\beta FF_0}{4F^2+\frac{a^4-1}{a^4}\gamma ^{-2}}.
\end{equation}

$0\leq \xi \leq d;$

\begin{eqnarray}
E_z^{^{\prime }}(\xi ) &=&\widetilde{E}_0\frac{n_b}{n_0}\frac{a\beta }{%
\lambda _{+}^2-\lambda _{-}^2}\left\{ \frac 1{\lambda _{+}}\left( \lambda
_{+}^2+\frac{a^4-1}{a^4}-4\beta ^2\gamma ^4F^2\right) \sin \left( \frac{%
\lambda _{+}}ak_p(d-\xi )\right) \right. - \\
&&\left. -\frac 1{\lambda _{-}}\left( \lambda _{-}^2+\frac{a^4-1}{a^4}%
-4\beta ^2\gamma ^4F^2\right) \sin \left( \frac{\lambda _{-}}ak_p(d-\xi
)\right) \right\} ,  \nonumber
\end{eqnarray}

\begin{eqnarray}
{\cal E}^{+}(\xi ) &=&i\widetilde{E}_0\frac{n_b}{n_0}\frac{\beta \sqrt{a^4-1}%
}a\left\{ \frac{\Delta F}{4F^2+\frac{a^4-1}{a^4}\gamma ^{-2}}k_p(d-2\xi
)-\right. \\
&&-\frac{\beta \gamma ^2}{\lambda _{+}^2-\lambda _{-}^2}\left[ \frac{\lambda
_{+}^2-2\beta \gamma ^2a\Delta F}{\lambda _{+}^3}\sin \left( \frac{\lambda
_{+}}ak_p(d-\xi )\right) -\frac{\lambda _{-}^2-2\beta \gamma ^2a\Delta F}{%
\lambda _{-}^3}\sin \left( \frac{\lambda _{-}}ak_p(d-\xi )\right) +\right. 
\nonumber \\
&&\left. \left. +i\left( a\Delta -2\beta \gamma ^2F\right) \left[ \frac
1{\lambda _{+}^2}\left( 1-\cos \left( \frac{\lambda _{+}}ak_p(d-\xi )\right)
\right) -\frac 1{\lambda _{-}^2}\left( 1-\cos \left( \frac{\lambda _{-}}%
ak_p(d-\xi )\right) \right) \right] \right] \right\} ,  \nonumber
\end{eqnarray}

\begin{eqnarray}
{\cal B}^{+}(\xi ) &=&\widetilde{E}_0\frac{n_b}{n_0}\frac{\beta \sqrt{a^4-1}%
}{a^2}\left\{ -\frac{FF_0}{4F^2+\frac{a^4-1}{a^4}\gamma ^{-2}}k_p(d-2\xi
)+\right. \\
&&\ +\frac{a\gamma ^2}{\lambda _{+}^2-\lambda _{-}^2}\left[ \frac{\lambda
_{+}^2-2\beta ^2\gamma ^2FF_0}{\lambda _{+}^3}\sin \left( \frac{\lambda _{+}}%
ak_p(d-\xi )\right) -\frac{\lambda _{-}^2-2\beta ^2\gamma ^2FF_0}{\lambda
_{-}^3}\sin \left( \frac{\lambda _{-}}ak_p(d-\xi )\right) -\right.  \nonumber
\\
&&\ \left. \left. -i\beta \left( 2\gamma ^2F-F_0\right) \left[ \frac
1{\lambda _{+}^2}\left( 1-\cos \left( \frac{\lambda _{+}}ak_p(d-\xi )\right)
\right) -\frac 1{\lambda _{-}^2}\left( 1-\cos \left( \frac{\lambda _{-}}%
ak_p(d-\xi )\right) \right) \right] \right] \right\} .  \nonumber
\end{eqnarray}

$\xi <0;$

\begin{eqnarray}
E_z^{^{\prime }}(\xi ) &=&\widetilde{E}_0\frac{n_b}{n_0}\frac{a\beta }{%
\lambda _{+}^2-\lambda _{-}^2}\left\{ \frac 1{\lambda _{+}}\left( \lambda
_{+}^2+\frac{a^4-1}{a^4}-4\beta ^2\gamma ^4F^2\right) \left[ \sin \left( 
\frac{\lambda _{+}}ak_p(d-\xi )\right) +\sin \left( \frac{\lambda _{+}}%
ak_p\xi \right) \right] -\right. \\
&&\left. -\frac 1{\lambda _{-}}\left( \lambda _{-}^2+\frac{a^4-1}{a^4}%
-4\beta ^2\gamma ^4F^2\right) \left[ \sin \left( \frac{\lambda _{-}}%
ak_p(d-\xi )\right) +\sin \left( \frac{\lambda _{-}}ak_p\xi \right) \right]
\right\} ,  \nonumber
\end{eqnarray}

\begin{eqnarray}
{\cal E}^{+}(\xi ) &=&i\widetilde{E}_0\frac{n_b}{n_0}\frac{\beta \sqrt{a^4-1}%
}a\left\{ \frac{\Delta F}{4F^2+\frac{a^4-1}{a^4}\gamma ^{-2}}(k_pd)-\right.
\\
&&-\frac{\beta \gamma ^2}{\lambda _{+}^2-\lambda _{-}^2}\left[ \frac{\lambda
_{+}^2-2\beta \gamma ^2a\Delta F}{\lambda _{+}^3}\left( \sin \left( \frac{%
\lambda _{+}}ak_p(d-\xi )\right) +\sin \left( \frac{\lambda _{+}}ak_p\xi
\right) \right) -\right.  \nonumber \\
&&-\frac{\lambda _{-}^2-2\beta \gamma ^2a\Delta F}{\lambda _{-}^3}\left(
\sin \left( \frac{\lambda _{-}}ak_p(d-\xi )\right) +\sin \left( \frac{%
\lambda _{-}}ak_p\xi \right) \right) +  \nonumber \\
&&+i\left( a\Delta -2\beta \gamma ^2F\right) \left[ \frac 1{\lambda
_{+}^2}\left( \cos \left( \frac{\lambda _{+}}ak_p\xi \right) -\cos \left( 
\frac{\lambda _{+}}ak_p(d-\xi )\right) \right) -\right.  \nonumber \\
&&\left. \left. \left. -\frac 1{\lambda _{-}^2}\left( \cos \left( \frac{%
\lambda _{-}}ak_p\xi \right) -\cos \left( \frac{\lambda _{-}}ak_p(d-\xi
)\right) \right) \right] \right] \right\} ,  \nonumber
\end{eqnarray}

\begin{eqnarray}
{\cal B}^{+}(\xi ) &=&\widetilde{E}_0\frac{n_b}{n_0}\frac{\beta \sqrt{a^4-1}%
}{a^2}\left\{ -\frac{FF_0}{4F^2+\frac{a^4-1}{a^4}\gamma ^{-2}}(k_pd)+\right.
\\
&&+\frac{a\gamma ^2}{\lambda _{+}^2-\lambda _{-}^2}\left[ \frac{\lambda
_{+}^2-2\beta ^2\gamma ^2FF_0}{\lambda _{+}^3}\left( \sin \left( \frac{%
\lambda _{+}}ak_p(d-\xi )\right) +\sin \left( \frac{\lambda _{+}}ak_p\xi
\right) \right) -\right.  \nonumber \\
&&-\frac{\lambda _{-}^2-2\beta ^2\gamma ^2FF_0}{\lambda _{-}^3}\left( \sin
\left( \frac{\lambda _{-}}ak_p(d-\xi )\right) +\sin \left( \frac{\lambda _{-}%
}ak_p\xi \right) \right) -  \nonumber \\
&&-i\beta \left( 2\gamma ^2F-F_0\right) \left[ \frac 1{\lambda _{+}^2}\left(
\cos \left( \frac{\lambda _{+}}ak_p\xi \right) -\cos \left( \frac{\lambda
_{+}}ak_p(d-\xi )\right) \right) -\right.  \nonumber \\
&&\left. \left. \left. -\frac 1{\lambda _{-}^2}\left( \cos \left( \frac{%
\lambda _{-}}ak_p\xi \right) -\cos \left( \frac{\lambda _{-}}ak_p(d-\xi
)\right) \right) \right] \right] \right\} .  \nonumber
\end{eqnarray}
Here we have adopted the notation $\widetilde{E}_0=mc\omega _p/e$.

\subsection{Region II}

In this range of values of bunch, plasma and pump wave parameters the
induced fields ahead of the bunch ($\xi >d$) coincide with the values of
induced EM fields in the region I and are given by the expressions
(56)-(58). The induced fields inside and behind the bunch are

$0\leq \xi \leq d;$

\begin{equation}
E_z^{^{\prime }}(\xi )=\widetilde{E}_0\frac{n_b}{n_0}\frac \beta {2\lambda
_0^2}\left[ a\frac{3\lambda _0^2-1}{\lambda _0}\sin \left( \frac{\lambda _0}%
ak_p(d-\xi )\right) +k_p(d-\xi )(1-\lambda _0^2)\cos \left( \frac{\lambda _0}%
ak_p(d-\xi )\right) \right] ,
\end{equation}

\begin{eqnarray}
{\cal E}^{+}(\xi ) &=&i\widetilde{E}_0\frac{n_b}{n_0}\frac{\beta \sqrt{a^4-1}%
}a\left\{ \frac{\Delta F}{4F^2+\frac{a^4-1}{a^4}\gamma ^{-2}}k_p(d-2\xi )+%
\frac{i\beta \gamma ^2}{\lambda _0^4}\left( a\Delta -2\beta \gamma
^2F\right) -\right. \\
&&\ -\frac{\beta \gamma ^2}{2\lambda _0^5}\left[ \left[ 6\beta \gamma
^2a\Delta F-\lambda _0^2+i\lambda _0^2\left( a\Delta -2\beta \gamma
^2F\right) \frac{k_p(d-\xi )}a\right] \sin \left( \frac{\lambda _0}%
ak_p(d-\xi )\right) +\right.  \nonumber \\
&&\ \left. \left. +\lambda _0\left[ \left( \lambda _0^2-2\beta \gamma
^2a\Delta F\right) \frac{k_p(d-\xi )}a+2i\left( a\Delta -2\beta \gamma
^2F\right) \right] \cos \left( \frac{\lambda _0}ak_p(d-\xi )\right) \right]
\right\} ,  \nonumber
\end{eqnarray}

\begin{eqnarray}
{\cal B}^{+}(\xi ) &=&\widetilde{E}_0\frac{n_b}{n_0}\frac{\beta \sqrt{a^4-1}%
}{a^2}\left\{ -\frac{FF_0}{4F^2+\frac{a^4-1}{a^4}\gamma ^{-2}}k_p(d-2\xi )+%
\frac{a\gamma ^2}{2\lambda _0^5}\left[ 2i\beta \lambda _0\left( 2\gamma
^2F-F_0\right) +\right. \right. \\
&&\ +\left[ 6\beta ^2\gamma ^2FF_0-\lambda _0^2-i\beta \lambda _0^2\left(
2\gamma ^2F-F_0\right) \frac{k_p(d-\xi )}a\right] \sin \left( \frac{\lambda
_0}ak_p(d-\xi )\right) +  \nonumber \\
&&\ \left. \left. +\lambda _0\left[ \left( \lambda _0^2-2\beta ^2\gamma
^2FF_0\right) \frac{k_p(d-\xi )}a-2i\beta \left( 2\gamma ^2F-F_0\right)
\right] \cos \left( \frac{\lambda _0}ak_p(d-\xi )\right) \right] \right\} . 
\nonumber
\end{eqnarray}

$\xi <0;$

\begin{eqnarray}
E_z^{^{\prime }}(\xi ) &=&\widetilde{E}_0\frac{n_b}{n_0}\frac \beta
{2\lambda _0^2}\left\{ a\frac{3\lambda _0^2-1}{\lambda _0}\left[ \sin \left( 
\frac{\lambda _0}ak_p(d-\xi )\right) +\sin \left( \frac{\lambda _0}ak_p\xi
\right) \right] +\right. \\
&&\left. +(1-\lambda _0^2)\left[ k_p(d-\xi )\cos \left( \frac{\lambda _0}%
ak_p(d-\xi )\right) +\left( k_p\xi \right) \cos \left( \frac{\lambda _0}%
ak_p\xi \right) \right] \right\} ,  \nonumber
\end{eqnarray}

\begin{eqnarray}
{\cal E}^{+}(\xi ) &=&i\widetilde{E}_0\frac{n_b}{n_0}\frac{\beta \sqrt{a^4-1}%
}a\left\{ \frac{\Delta Fk_pd}{4F^2+\frac{a^4-1}{a^4}\gamma ^{-2}}-\right. \\
&&-\frac{\beta \gamma ^2}{2\lambda _0^5}\left[ \left[ 6\beta \gamma
^2a\Delta F-\lambda _0^2+i\lambda _0^2\left( a\Delta -2\beta \gamma
^2F\right) \frac{k_p(d-\xi )}a\right] \sin \left( \frac{\lambda _0}%
ak_p(d-\xi )\right) +\right.  \nonumber \\
&&+\left[ 6\beta \gamma ^2a\Delta F-\lambda _0^2-i\lambda _0^2\left( a\Delta
-2\beta \gamma ^2F\right) \frac{k_p\xi }a\right] \sin \left( \frac{\lambda _0%
}ak_p\xi \right) +  \nonumber \\
&&+\lambda _0\left[ \left( \lambda _0^2-2\beta \gamma ^2a\Delta F\right) 
\frac{k_p(d-\xi )}a+2i\left( a\Delta -2\beta \gamma ^2F\right) \right] \cos
\left( \frac{\lambda _0}ak_p(d-\xi )\right) +  \nonumber \\
&&\left. \left. +\lambda _0\left[ \left( \lambda _0^2-2\beta \gamma
^2a\Delta F\right) \frac{k_p\xi }a-2i\left( a\Delta -2\beta \gamma
^2F\right) \right] \cos \left( \frac{\lambda _0}ak_p\xi \right) \right]
\right\} ,  \nonumber
\end{eqnarray}

\begin{eqnarray}
{\cal B}^{+}(\xi ) &=&\widetilde{E}_0\frac{n_b}{n_0}\frac{\beta \sqrt{a^4-1}%
}{a^2}\left\{ -\frac{FF_0}{4F^2+\frac{a^4-1}{a^4}\gamma ^{-2}}(k_pd)+\right.
\\
&&+\frac{a\gamma ^2}{2\lambda _0^5}\left[ \left[ 6\beta ^2\gamma
^2FF_0-\lambda _0^2-i\beta \lambda _0^2\left( 2\gamma ^2F-F_0\right) \frac{%
k_p(d-\xi )}a\right] \sin \left( \frac{\lambda _0}ak_p(d-\xi )\right)
+\right.  \nonumber \\
&&+\left[ 6\beta ^2\gamma ^2FF_0-\lambda _0^2+i\beta \lambda _0^2\left(
2\gamma ^2F-F_0\right) \frac{k_p\xi }a\right] \sin \left( \frac{\lambda _0}%
ak_p\xi \right) +  \nonumber \\
&&+\lambda _0\left[ \left( \lambda _0^2-2\beta ^2\gamma ^2FF_0\right) \frac{%
k_p(d-\xi )}a-2i\beta \left( 2\gamma ^2F-F_0\right) \right] \cos \left( 
\frac{\lambda _0}ak_p(d-\xi )\right) +  \nonumber \\
&&\left. \left. +\lambda _0\left[ \left( \lambda _0^2-2\beta ^2\gamma
^2FF_0\right) \frac{k_p\xi }a+2i\beta \left( 2\gamma ^2F-F_0\right) \right]
\cos \left( \frac{\lambda _0}ak_p\xi \right) \right] \right\} ,  \nonumber
\end{eqnarray}
where $\lambda _0$ is determined by the Eq. (45).

\subsection{Region III}

In this region the Green's functions are determined by Eqs. (52)-(54). The
expression for longitudinal electric fields over the space ($-\infty <\xi
<+\infty $) is obtained after the substitution of Eqs. (52)-(54) into the
Eq. (22) and appropriate integrations:

\begin{eqnarray}
E_z^{^{\prime }}(\xi ) &=&\widetilde{E}_0\frac{n_b}{n_0}\frac{a\beta }{%
4\left( \alpha ^2+\delta ^2\right) ^2}\left\{ -\frac 1\alpha \left( \alpha
^2+\delta ^2-\frac{a^4-1}{a^4}+4\beta ^2\gamma ^4F^2\right) \times \right. \\
&&\ \times \left[ \exp \left( -\frac \delta ak_p|\xi |\right) \sin \left(
\frac \alpha ak_p|\xi |\right) -\exp \left( -\frac \delta ak_p|\xi
-d|\right) \sin \left( \frac \alpha ak_p|\xi -d|\right) \right] +  \nonumber
\\
&&\ +\frac 1\delta \left( \alpha ^2+\delta ^2+\frac{a^4-1}{a^4}-4\beta
^2\gamma ^4F^2\right) \times  \nonumber \\
&&\ \left. \times \left[ \exp \left( -\frac \delta ak_p|\xi |\right) \cos
\left( \frac \alpha ak_p\xi \right) -\exp \left( -\frac \delta ak_p|\xi
-d|\right) \cos \left( \frac \alpha ak_p(\xi -d)\right) \right] \right\} . 
\nonumber
\end{eqnarray}

The expressions for induced transverse electromagnetic fields ahead of ($\xi
>d$), inside ($0\leqslant \xi \leqslant d$) and behind ($\xi <0$) the bunch
are obtained in similar manner:

$\xi >d;$

\begin{eqnarray}
{\cal E}^{+}(\xi ) &=&-i\widetilde{E}_0\frac{n_b}{n_0}(k_pd)\frac{\sqrt{a^4-1%
}}a\frac{\beta \Delta F}{4F^2+\frac{a^4-1}{a^4}\gamma ^{-2}}+\widetilde{E}_0%
\frac{n_b}{n_0}\frac{\beta ^2\gamma ^2}{4\alpha \delta \left( \alpha
^2+\delta ^2\right) ^2}\frac{\sqrt{a^4-1}}a\times \\
&&\ \times \left\{ \left[ 2\alpha \delta \left( a\Delta -2\beta \gamma
^2F\right) -\frac{i\alpha }{\alpha ^2+\delta ^2}\left( \left( \alpha
^2+\delta ^2\right) ^2+2\beta \gamma ^2Fa\Delta \left( 3\delta ^2-\alpha
^2\right) \right) \right] \times \right.  \nonumber \\
&&\ \times \left[ \exp \left( -\frac \delta ak_p(\xi -d)\right) \cos \left(
\frac \alpha ak_p(\xi -d)\right) -\exp \left( -\frac \delta ak_p\xi \right)
\cos \left( \frac \alpha ak_p\xi \right) \right] +  \nonumber \\
&&\ +\left[ \left( a\Delta -2\beta \gamma ^2F\right) \left( \alpha ^2-\delta
^2\right) +\frac{i\delta }{\alpha ^2+\delta ^2}\left( \left( \alpha
^2+\delta ^2\right) ^2-2\beta \gamma ^2Fa\Delta \left( 3\alpha ^2-\delta
^2\right) \right) \right] \times  \nonumber \\
&&\ \left. \times \left[ \exp \left( -\frac \delta ak_p\xi \right) \sin
\left( \frac \alpha ak_p\xi \right) -\exp \left( -\frac \delta ak_p(\xi
-d)\right) \sin \left( \frac \alpha ak_p(\xi -d)\right) \right] \right\} , 
\nonumber
\end{eqnarray}

\begin{eqnarray}
{\cal B}^{+}(\xi ) &=&\widetilde{E}_0\frac{n_b}{n_0}\left( k_pd\right) \frac{%
\sqrt{a^4-1}}{a^2}\frac{\beta FF_0}{4F^2+\frac{a^4-1}{a^4}\gamma ^{-2}}+%
\widetilde{E}_0\frac{n_b}{n_0}\frac{\sqrt{a^4-1}}a\frac{\beta \gamma ^2}{%
4\alpha \delta \left( \alpha ^2+\delta ^2\right) ^2}\times \\
&&\ \times \left\{ -\alpha \left[ \frac{\left( \alpha ^2+\delta ^2\right)
^2+2\beta ^2\gamma ^2\left( 3\delta ^2-\alpha ^2\right) FF_0}{\alpha
^2+\delta ^2}+2i\beta \delta \left( F_0-2\gamma ^2F\right) \right] \times
\right.  \nonumber \\
&&\ \times \left[ \exp \left( -\frac \delta ak_p(\xi -d)\right) \cos \left(
\frac \alpha ak_p(\xi -d)\right) -\exp \left( -\frac \delta ak_p\xi \right)
\cos \left( \frac \alpha ak_p\xi \right) \right] +  \nonumber \\
&&\ +\left[ \delta \frac{\left( \alpha ^2+\delta ^2\right) ^2-2\beta
^2\gamma ^2\left( 3\alpha ^2-\delta ^2\right) FF_0}{\alpha ^2+\delta ^2}%
+i\beta \left( \delta ^2-\alpha ^2\right) \left( F_0-2\gamma ^2F\right)
\right] \times  \nonumber \\
&&\ \left. \times \left[ \exp \left( -\frac \delta ak_p\xi \right) \sin
\left( \frac \alpha ak_p\xi \right) -\exp \left( -\frac \delta ak_p(\xi
-d)\right) \sin \left( \frac \alpha ak_p(\xi -d)\right) \right] \right\} . 
\nonumber
\end{eqnarray}

$0\leqslant \xi \leqslant d;$

\begin{eqnarray}
{\cal E}^{+}(\xi ) &=&i\widetilde{E}_0\frac{n_b}{n_0}\left( k_p(d-2\xi
)\right) \frac{\sqrt{a^4-1}}a\frac{\beta \Delta F}{4F^2+\frac{a^4-1}{a^4}%
\gamma ^{-2}}-\widetilde{E}_0\frac{n_b}{n_0}\frac{\beta ^2\gamma ^2}{4\alpha
\delta \left( \alpha ^2+\delta ^2\right) ^2}\frac{\sqrt{a^4-1}}a\times \\
&&\ \times \left\{ 2\alpha \delta \left( a\Delta -2\beta \gamma ^2F\right)
\left[ \exp \left( -\frac \delta ak_p\xi \right) \cos \left( \frac \alpha
ak_p\xi \right) -\exp \left( -\frac \delta ak_p(d-\xi )\right) \cos \left(
\frac \alpha ak_p(d-\xi )\right) \right] +\right.  \nonumber \\
&&\ +\left( a\Delta -2\beta \gamma ^2F\right) \left( \delta ^2-\alpha
^2\right) \left[ \exp \left( -\frac \delta ak_p\xi \right) \sin \left( \frac
\alpha ak_p\xi \right) -\exp \left( -\frac \delta ak_p(d-\xi )\right) \sin
\left( \frac \alpha ak_p(d-\xi )\right) \right] -  \nonumber \\
&&\ -\frac{i\alpha }{\alpha ^2+\delta ^2}\left[ \left( \alpha ^2+\delta
^2\right) ^2-2\beta \gamma ^2Fa\Delta \left( \alpha ^2-3\delta ^2\right)
\right] \times  \nonumber \\
&&\ \times \left[ \exp \left( -\frac \delta ak_p\xi \right) \cos \left(
\frac \alpha ak_p\xi \right) +\exp \left( -\frac \delta ak_p(d-\xi )\right)
\cos \left( \frac \alpha ak_p(d-\xi )\right) -2\right] -  \nonumber \\
&&\ -\frac{i\delta }{\alpha ^2+\delta ^2}\left[ \left( \alpha ^2+\delta
^2\right) ^2-2\beta \gamma ^2Fa\Delta \left( 3\alpha ^2-\delta ^2\right)
\right] \times  \nonumber \\
&&\ \left. \times \left[ \exp \left( -\frac \delta ak_p\xi \right) \sin
\left( \frac \alpha ak_p\xi \right) +\exp \left( -\frac \delta ak_p(d-\xi
)\right) \sin \left( \frac \alpha ak_p(d-\xi )\right) \right] \right\} , 
\nonumber
\end{eqnarray}

\begin{eqnarray}
{\cal B}^{+}(\xi ) &=&-\widetilde{E}_0\frac{n_b}{n_0}\frac{\sqrt{a^4-1}}{a^2}%
\left( k_p(d-2\xi )\right) \frac{\beta FF_0}{4F^2+\frac{a^4-1}{a^4}\gamma
^{-2}}+\widetilde{E}_0\frac{n_b}{n_0}\frac{\sqrt{a^4-1}}a\frac{\beta \gamma
^2}{4\alpha \delta \left( \alpha ^2+\delta ^2\right) ^2}\times \\
&&\ \times \left\{ -\alpha \left[ \frac{\left( \alpha ^2+\delta ^2\right)
^2+2\beta ^2\gamma ^2\left( 3\delta ^2-\alpha ^2\right) FF_0}{\alpha
^2+\delta ^2}+2i\beta \delta \left( F_0-2\gamma ^2F\right) \right] \left[
1-\exp \left( -\frac \delta ak_p\xi \right) \cos \left( \frac \alpha ak_p\xi
\right) \right] +\right.  \nonumber \\
&&\ +\alpha \left[ \frac{\left( \alpha ^2+\delta ^2\right) ^2+2\beta
^2\gamma ^2\left( 3\delta ^2-\alpha ^2\right) FF_0}{\alpha ^2+\delta ^2}%
-2i\beta \delta \left( F_0-2\gamma ^2F\right) \right] \left[ 1-\exp \left(
-\frac \delta ak_p(d-\xi )\right) \cos \left( \frac \alpha ak_p(d-\xi
)\right) \right] +  \nonumber \\
&&\ +\left[ \delta \frac{\left( \alpha ^2+\delta ^2\right) ^2-2\beta
^2\gamma ^2\left( 3\alpha ^2-\delta ^2\right) FF_0}{\alpha ^2+\delta ^2}%
+i\beta \left( \delta ^2-\alpha ^2\right) \left( F_0-2\gamma ^2F\right)
\right] \exp \left( -\frac \delta ak_p\xi \right) \sin \left( \frac \alpha
ak_p\xi \right) -  \nonumber \\
&&\ \left. -\left[ \delta \frac{\left( \alpha ^2+\delta ^2\right) ^2-2\beta
^2\gamma ^2\left( 3\alpha ^2-\delta ^2\right) FF_0}{\alpha ^2+\delta ^2}%
-i\beta \left( \delta ^2-\alpha ^2\right) \left( F_0-2\gamma ^2F\right)
\right] \exp \left( -\frac \delta ak_p(d-\xi )\right) \sin \left( \frac
\alpha ak_p(d-\xi )\right) \right\} .  \nonumber
\end{eqnarray}

$\xi <0;$

\begin{eqnarray}
{\cal E}^{+}(\xi ) &=&i\widetilde{E}_0\frac{n_b}{n_0}(k_pd)\frac{\sqrt{a^4-1}%
}a\frac{\beta \Delta F}{4F^2+\frac{a^4-1}{a^4}\gamma ^{-2}}-\widetilde{E}_0%
\frac{n_b}{n_0}\frac{\beta ^2\gamma ^2}{4\alpha \delta \left( \alpha
^2+\delta ^2\right) ^2}\frac{\sqrt{a^4-1}}a\times \\
&&\times \left\{ \alpha \left[ 2\delta \left( a\Delta -2\beta \gamma
^2F\right) +i\frac{\left( \alpha ^2+\delta ^2\right) ^2-2\beta \gamma
^2Fa\Delta \left( \alpha ^2-3\delta ^2\right) }{\alpha ^2+\delta ^2}\right]
\times \right.  \nonumber \\
&&\times \left[ \exp \left( \frac \delta ak_p\xi \right) \cos \left( \frac
\alpha ak_p\xi \right) -\exp \left( -\frac \delta ak_p(d-\xi )\right) \cos
\left( \frac \alpha ak_p(d-\xi )\right) \right] +  \nonumber \\
&&+\left[ \left( a\Delta -2\beta \gamma ^2F\right) \left( \alpha ^2-\delta
^2\right) +i\delta \frac{2\beta \gamma ^2Fa\Delta \left( 3\alpha ^2-\delta
^2\right) -\left( \alpha ^2+\delta ^2\right) ^2}{\alpha ^2+\delta ^2}\right]
\times  \nonumber \\
&&\left. \times \left[ \exp \left( \frac \delta ak_p\xi \right) \sin \left(
\frac \alpha ak_p\xi \right) +\exp \left( -\frac \delta ak_p(d-\xi )\right)
\sin \left( \frac \alpha ak_p(d-\xi )\right) \right] \right\} ,  \nonumber
\end{eqnarray}

\begin{eqnarray}
{\cal B}^{+}(\xi ) &=&-\widetilde{E}_0\frac{n_b}{n_0}\frac{\sqrt{a^4-1}}{a^2}%
(k_pd)\frac{\beta FF_0}{4F^2+\frac{a^4-1}{a^4}\gamma ^{-2}}+\widetilde{E}_0%
\frac{n_b}{n_0}\frac{\sqrt{a^4-1}}a\frac{\beta \gamma ^2}{4\alpha \delta
\left( \alpha ^2+\delta ^2\right) ^2}\times \\
&&\ \times \left\{ \alpha \left[ \frac{\left( \alpha ^2+\delta ^2\right)
^2+2\beta ^2\gamma ^2\left( 3\delta ^2-\alpha ^2\right) FF_0}{\alpha
^2+\delta ^2}-2i\beta \delta \left( F_0-2\gamma ^2F\right) \right] \times
\right.  \nonumber \\
&&\ \times \left[ \exp \left( \frac \delta ak_p\xi \right) \cos \left( \frac
\alpha ak_p\xi \right) -\exp \left( -\frac \delta ak_p(d-\xi )\right) \cos
\left( \frac \alpha ak_p(d-\xi )\right) \right] -  \nonumber \\
&&\ -\left[ \delta \frac{\left( \alpha ^2+\delta ^2\right) ^2-2\beta
^2\gamma ^2\left( 3\alpha ^2-\delta ^2\right) FF_0}{\alpha ^2+\delta ^2}%
-i\beta \left( \delta ^2-\alpha ^2\right) \left( F_0-2\gamma ^2F\right)
\right] \times  \nonumber \\
&&\ \left. \times \left[ \exp \left( \frac \delta ak_p\xi \right) \sin
\left( \frac \alpha ak_p\xi \right) +\exp \left( -\frac \delta ak_p(d-\xi
)\right) \sin \left( \frac \alpha ak_p(d-\xi )\right) \right] \right\} . 
\nonumber
\end{eqnarray}

As it follows from the expression (56), in regions I and II the longitudinal
field ahead of bunch ($\xi >d$) is missing. The transverse fields ahead of
the bunch are independent of $\xi $ (i.e., are not modulated), are
circularly polarized and proportional to the function $F$ and thickness of
the bunch $d$. Note that $F$ is proportional to the difference between the
group velocity of induced transverse fields ($v_g=k_0c^2/\omega _0$) and the
bunch velocity. If the group velocity of the transverse wave coincides with
the bunch velocity ($F=0$), then the transverse fields ahead of the bunch
are absent. Besides, the magnetic field ahead of the bunch is proportional
to $F_0$ and is absent in the long-wave (quasi-stationary) limit ($k_0=0$ or 
$F_0=0$).

The transverse fields ahead of the bunch arise on account of the fact that
the CPEM wave has a phase velocity ($v_\varphi =\omega _0/k_0>c$) exceeding
the bunch velocity at any values of parameters of plasma and of the CPEM
wave. Hence, some part of perturbations caused in plasma by the CPEM wave
and bunch may have phase velocity higher than the bunch velocity and be even
ahead of it.

Inside and behind of bunch for values of parameters in regions I, II, III,
as well as ahead of it in the region III (see the Eqs. (59), (61), (63),
(64), (66), (67), (69), (70), (72)-(77)) the transverse fields are modulated
and are circularly polarized. Though the polarization vector of the
transverse wave describes a circle, its radius depends on the distance to
the bunch $\xi $. Indeed, as it follows from Eqs. (29)-(31)

\begin{equation}
E_x^{\prime 2}+E_y^{\prime 2}=E_r^2(\xi )+E_i^2(\xi )=E_{\max }^2(\xi ),
\end{equation}

\begin{equation}
B_x^{\prime 2}+B_y^{\prime 2}=B_r^2(\xi )+B_i^2(\xi )=B_{\max }^2(\xi ).
\end{equation}
The amplitudes of electric ($E_{\max }(\xi )$) and magnetic ($B_{\max }(\xi
) $) fields are generally dependent on $\xi $. The modulated transverse
waves in plasma arise in consequence of the excitation of two type waves,
having frequencies $\omega _L\lambda _{\pm }$ and wave vectors ($\omega
_L/u)\lambda _{\pm }$. Owing to the interaction of the pump wave with these
induced waves, there arise waves with combined frequencies $\omega _0-\omega
_L\lambda _{\pm }$, $\omega _0+\omega _L\lambda _{\pm }$ and combined wave
vectors $k_0-(\omega _L/u)\lambda _{\pm }$, $k_0+(\omega _L/u)\lambda _{\pm
} $, the interference between which gives a modulated wave.

Here $E_r(\xi )$ or $E_i(\xi )$ serve as a carrier wave if $\omega _0>\omega
_L\lambda _{\pm }$ (in the dimensionless form $a\Delta >\lambda _{\pm }$)
and $k_0>(\omega _L/u)\lambda _{\pm }$ (in the dimensionless form $\beta 
\sqrt{a^2\Delta ^2-1}>\lambda _{\pm }$). Otherwise in expressions (29) and
(30) the part of the carrier wave will play the functions $\cos \zeta $ and $%
\sin \zeta $.

Below we shall analyze the obtained expressions in practically important
case of the following values of parameters: $n_0\lesssim 10^{17}{\tt cm}%
^{-3} $, ($\omega _p\lesssim 2\times 10^{13}\sec ^{-1}$), $W_L\simeq 10^{18}$%
W/cm$^2\div 10^{20}$W/cm$^2$ ($W_L=cE_0^2/4\pi $ is the intensity of CPEM
wave), $\omega _0\simeq 10^{15}\sec ^{-1}$, $\gamma \simeq 10\div 10^3$. At
the variation of CPEM wave intensity in the range from $10^{18}$W/cm$^2$ to $%
10^{20}$W/cm$^2$ the parameter $a$ $\,$is changed in the interval $a\simeq
1.02\div 2$. For the parameter $\Delta $ we have: $\Delta \gtrsim 50$.

In the region I $\gamma \lesssim \Delta $ or $\gamma \gtrsim \Delta $ (the
Eq. (40)). It is comparatively easy to analyze the Eqs. (56)-(64)\ in two
particular cases of $a\Delta \gg \gamma \gg 1$ and $1\ll a\Delta \ll \gamma $
respectively. In the first case we obtain from Eqs. (34) and (35):

\begin{equation}
4\gamma ^4F^2\simeq a^2\Delta ^2\left( 1-\frac{2\gamma ^2}{a^2\Delta ^2}%
\right) \gg \gamma ^2\gg 1,
\end{equation}

\begin{equation}
\lambda _{-}^2\simeq 1,\quad \lambda _{+}^2\simeq 4\gamma ^4F^2\simeq
a^2\Delta ^2\gg \lambda _{-}^2.
\end{equation}

The transverse electric and magnetic fields ahead of the bunch are on the
order of magnitude $|{\cal E}^{+}|\simeq \left| {\cal B}^{+}\right| \simeq 
\widetilde{E}_0(n_b/n_0)(k_pd)\gamma ^2/2$ when $a\gtrsim 1$. Inside and
behind the bunch the amplitude of induced waves with frequency $\omega
_L\lambda _{+}$ is much less than the one for waves with frequency $\omega
_L\lambda _{-}$. By the order of magnitude the amplitude of oscillations
with frequency $\omega _L\lambda _{-}$ in the longitudinal waves (Eqs. (59)
and (62) is $\widetilde{E}_0(n_b/n_0)a$, i.e., is $\gamma ^2(k_pd)/2a$ times
as less as the amplitude of transverse waves ahead of the bunch. For $%
0\leqslant \xi \leqslant d$ and $\xi <0$ the highest contributions to
expressions (60), (61), (63), (64) are made by the first and third terms,
the amplitude of the third term being approximately equal to $\widetilde{E}%
_0(n_b/n_0)\gamma ^2a$ (i.e., $\gamma ^2$ times as much as the amplitude of
longitudinal waves). Hence, in the region I $E_i(\xi )\gg E_r(\xi )\simeq 0$%
, $B_r(\xi )\gg B_i(\xi )\simeq 0$ and $E_x^{^{\prime }}\cong -E_i(\xi )\sin
\zeta $, $E_y^{^{\prime }}\cong E_i(\xi )\cos \zeta $, $B_x^{^{\prime
}}\cong B_r(\xi )\cos \zeta $, $B_y^{^{\prime }}\cong B_i(\xi )\sin \zeta $.
Besides, the oscillating terms in Eqs. (59)-(64) describe the waves, the
length of which increases with the intensity of CPEM wave. Third, when the
condition $\pi d<a\lambda _p$ is met (where $\lambda _p=2\pi /k_p$ is the
wavelength of excited longitudinal waves in the absence of the CPEM wave),
the oscillating terms in expressions (60), (61), (63), (64) exceed the first
terms for narrow bunches. In case of wide bunches ($\pi d>a\lambda _p$) the
first term in Eqs. (63) and (64) exceeds the third term. Inside the wide
bunch the analogous conclusion is valid close to the bunch boundaries when $%
\pi \left| d/2-\xi \right| >a\lambda _p/2$. Inside the wide bunch, in the
vicinity of its center, $\pi \left| d/2-\xi \right| <a\lambda _p/2$, the
main contribution to Eqs. (60) and (61) is made by the third terms.

Behind the bunch, the amplitude of longitudinal and transverse waves is
proportional to $2\sin (\pi d/a\lambda _p)$ and is maximum at $%
d=(n-1/2)a\lambda _p$, where $n=1,2,...$. When the condition $d=(a\lambda
_p)n$ is fulfilled, the waves behind the bunch are not excited.

In Figs. 4-10 are shown the results of numerical calculations made for
induced fields using Eqs. (56)-(64). The figures illustrate all the features
discussed earlier. Thus, in the presence of a CPEM wave in the plasma, the
one-dimensional bunch excites the waves with the wavelengths increasing as
the intensity of CPEM wave. The dependence of longitudinal wave amplitude on
the energy of bunch (on the relativistic factor) is weak and grows with the
intensity of CPEM wave, whereas the amplitude of the transverse wave is $%
\gamma ^2$ times as large as that of longitudinal wave. One can assert that
at $\gamma \gg 1$ the induced wave is almost a transverse one.

Now consider another limiting case with $\gamma \gg a\Delta \gg $1. Here
instead of Eqs. (80) and (81) we find

\begin{equation}
4\gamma ^4F^2\simeq \gamma ^2\left( \frac \gamma {a\Delta }\right) ^2\gg 1,
\end{equation}

\begin{equation}
\lambda _{-}^2\simeq 1,\quad \lambda _{+}^2\simeq \gamma ^2\left( \frac
\gamma {a\Delta }\right) ^2\gg \lambda _{-}^2.
\end{equation}

When $a\gtrsim 1,$ the transverse fields ahead of the bunch are by the order
of magnitude: $\left| {\cal E}^{+}\right| \simeq \left| {\cal B}^{+}\right|
\simeq \widetilde{E}_0(n_b/n_0)(k_pd)a^2\Delta ^2/2$. Inside and behind the
bunch the amplitude of waves with frequency $\omega _L\lambda _{+}$ is again
much less than the amplitude of waves with frequency $\omega _L\lambda _{-}$%
. The amplitude of oscillations with frequency $\omega _L\lambda _{-}$ in
the longitudinal waves (Eqs. (59) and (62)) does not change and is nearly
equal to $\widetilde{E}_0(n_b/n_0)a$. In this case the magnitude of the
longitudinal wave is by $(k_pd)a\Delta ^2/2$ times less than the transverse
wave amplitude ahead of the bunch. Inside and behind the bunch the primary
role in the transverse waves is played again by the first and third terms.
The amplitude of the third terms is $\widetilde{E}_0(n_b/n_0)a(a\Delta )^2$
(i.e., by $(a\Delta )^2$ times larger than the amplitude of longitudinal
waves). Note that in the limit of very large values of $\gamma $ $(\gamma
\gg a\Delta )$ the transverse fields are independent of it, but the
dependence on intensity of CPEM wave becomes more salient. All the remaining
features mentioned above for the case of $\gamma \ll a\Delta $ are still
valid here.

As we saw in Sec. III, $\gamma \sim a\Delta $ in the second region of values
of $a$, $\Delta $, $\gamma $. In this case the function $F$ in Eqs.
(56)-(58), (65)-(70) takes on the values $F_{\pm }$ (see the expression
(38)), the region II being determined by expressions (36) for $\gamma \gg 1$
and an arbitrary value of parameter $a$, and by expressions (37) for $\gamma
\gg 1$ and $1<a<a_0(\gamma )$ (where $a_0(\gamma )$ is determined by the
expression (39)). The values $F=\pm F_{+}$ correspond to Eq. (36), and the
values $F=\pm F_{-}$ correspond to Eq. (37). Respectively, in the expression
(45) one is to take the sign ''$+$'' if $F=\pm \,F_{+}$, and the sign ''$-$%
'' if $F=\pm F_{-}$. From expression (45) for values $\gamma \gg 1$, $\Delta
\gg 1$ we have $\lambda _0\simeq \gamma ^{1/2}\gg 1$ if $a\gtrsim
1+1/4\gamma ^2$, and $\lambda _0\lesssim 1$ if $1<a<a_0(\gamma )\simeq
1+1/16\gamma ^2$. In the first case the wavelength of the waves excited
inside and behind the bunch is decreased as $\gamma $ increases and is much
less than the wavelength in the second case, when that is on the order of $%
a\lambda _p$.

As it follows from Eqs. (65)-(70), at the distance from the bunch the
amplitudes of waves to be excited increase proportional to the distance to
bunch and beginning from some value $\xi =\xi _0$ the fields may be of the
same order of magnitude as the amplitude of CPEM wave. Surely, in case of $%
\xi >\xi _0$ the Eqs. (68)-(70) are not adequate for the description of
induced fields and for their treatment one is to avail of the nonlinear
approach.

The obtained results for region II were further analyzed numerically. In
Figs. 11-14 the coordinate dependence of induced longitudinal and transverse
fields for the following values of CPEM wave, plasma and bunch parameters is
shown: $n_b=10^{14}$cm$^{-3}$ (Fig. 11), $n_b=10^{13}$cm$^{-3}$ (Figs.
12-14), $n_0=10^{17}$cm$^{-3}$, $\omega _0=3.77\times 10^{15}$sec$^{-1}$, $%
\gamma =210$, $d=2k_p^{-1}$, $\,k_p^{-1}=1.7\times 10^{-3}$cm. The strength
of CPEM wave field is varied from the value $E_0=8.5\times 10^8$V/cm to $%
E_0=1.85\times 10^{10}$V/cm. The $\xi $-dependence of $B_{\max }$ in Figs.
11-14 is not shown, since the difference of $B_{\max }$ from $E_{\max }(\xi
) $ is insignificant as it follows from Eqs. (66), (67) and (69), (70) for $%
a\Delta \gg 1$, $\gamma \gg 1$. From Fig. 12 one can estimate the distance $%
\xi _0$, at which the amplitude of transverse waves is of the order of CPEM
wave amplitude and the linear approach is of no sense any more. From Fig. 12
one gets the following estimate: $\xi _0\simeq 60k_p^{-1}$. So, the linear
treatment is valid at distances of several wavelengths $(2\pi /k_p)$.

In the region III the fields are described by Eqs. (71)-(77). At large
distances from the bunch $\left| \xi \right| \gtrsim a(\delta k_p)^{-1}$, $%
\left| \xi -d\right| \gtrsim a(\delta k_p)^{-1}$, the longitudinal waves are
attenuated, and transverse waves are described by the first terms in Eqs.
(72), (73) and (76), (77) that are independent of $\xi $ (i.e., are not
modulated). The $\xi $-dependence of transverse and longitudinal waves is
given in Figs. 15-21. It follows from these figures that the longitudinal
field and amplitudes $E_r$, $E_i$, $B_r$ are antisymmetric with respect to
the bunch center ($\xi =d/2$ plane), and the amplitude $B_i$ is symmetric
with respect to it. Besides, at large distances the phase of transverse
waves ahead of the bunch differs from the phase of transverse wave behind
the bunch by $\pi $, and the transverse oscillations behind and ahead of the
bunch are in the counterphase.

\section{Summary}

The purpose of this work was to investigate the excitation of EM wake waves
in electron plasma by one-dimensional bunch of charged particles in the
presence of intense CPEM wave. The interaction of the pump wave with a
plasma was described by means of Maxwell equations and relativistic
nonlinear hydrodynamic equations of plasma. Then, we have considered small
perturbations of plasma due to the presence of one-dimensional bunch. A
general expressions obtained for EM wake field components was analyzed in
three particular ranges of parameters of the pump wave, bunch and plasma. In
the first range of parameters the amplitude of transverse components of
induced waves is shown to grow as the bunch energy and after some definite
value of the relativistic factor of the bunch to be almost independent of
the energy and increase proportional to the intensity and frequency of the
CPEM wave. The dependence of the longitudinal component of induced waves on
the relativistic factor of the bunch is weak. Its amplitude and wavelength
grow as the intensity of CPEM wave. In the second range of parameters the
amplitude of the wave excited by the bunch is a linear function of the
distance to the bunch. In the third range of parameters the longitudinal
component of induced fields is localized near the bunch boundaries and
exponentially decreases with increasing distance from these boundaries. The
amplitudes of transverse components of induced waves reach a constant value
with the distance from the bunch boundaries. In general, the transverse
waves in all three mentioned ranges of values of the CPEM wave, plasma and
bunch parameters are modulated and are circularly polarized with the same
type (right-hand or left-hand) of polarization as the pump wave.

In conclusion it is worthwhile to make two remarks. First, although at the
consideration of EM wake wave generation we took the external CPEM wave to
be strong, it was assumed monochromatic. The intensity of present-day
sources of monochromatic EM waves does not exceed $10^{16}$W/cm$^2$ and the
values of amplitudes used for numerical estimates and calculations are much
higher than this value. In the nearest future we plan to study the
excitation of EM wake waves by one-dimensional bunch in the presence of high
power laser pulses with the intensity $\symbol{126}10^{20}$W/cm$^2$. The
results of this study will be presented in a forthcoming work. Second, it
follows from Eqs. (10)-(14) that in the plasma there arise parametrically
coupled linear waves. In case of high bunch densities ($n_b\gtrsim n_0$) the
linear treatment is of nosense any more and it is necessary to examine
explicit system of nonlinear Eqs. (1)-(5). This issue will also be studied
in future work.

\vspace{1.0in}

\begin{center}
{\bf ACKNOWLEDGMENT}
\end{center}

The work has been supported by the International Science and Technology
Center under Project No. A-013.

\newpage\ 

\section{Appendix}

In Section II a set of inhomogeneous differential equations with constant
coefficients for induced variables ${\cal E}^{\pm },$ $B^{\pm },$ ${\cal V}%
^{\pm },$ $E_z^{^{\prime }}$ and $n^{^{\prime }}$ (see Eqs. (16)-(20)) has
been found. The solution of this set of equations was obtained with the help
of Green's function. Below we shall briefly discuss another method for
solution of the set of Eqs. (16)-(20), i.e., we shall obtain a differential
equation for one of the induced quantities, i.e., for the density of induced
charge $n^{^{\prime }}$. The equations for other variables will be obtained
in an analogous way.

Now consider the Eq. (20) and find the differential operator the action of
which on ${\cal B}^{+}+{\cal B}^{-}$ and ${\cal V}^{+}+{\cal V}^{-}$ give a
function of $n^{^{\prime }}$. Let us introduce the operators

\begin{equation}
\widehat{Q}^{\pm }=\frac 1{\gamma ^2}\frac{\partial ^2}{\partial \xi ^2}\mp
2i\left( k_0-\beta \frac{\omega _0}c\right) \frac \partial {\partial \xi }+%
\frac{\omega _L^2}{c^2}.  \tag{A.1}
\end{equation}
After successive action of operators $\widehat{Q}^{-}$ and $\widehat{Q}^{+}$
on both the sides of the Eq. (18) we obtain the equation that comprises only
the functions ${\cal V}^{\pm }(\xi )$ and $n^{^{\prime }}(\xi )$:

\begin{equation}
\widehat{Q}^{+}\widehat{Q}^{-}{\cal V}^{\pm }(\xi )=\frac{\omega _L^2}{c^2}%
\left[ \left( 1-\frac{\beta _e^2}2\right) \widehat{Q}^{\mp }\left( {\cal V}%
^{\pm }(\xi )\pm iv_e\frac{n^{^{\prime }}(\xi )}{n_0}\right) +\frac{\beta
_e^2}2\widehat{Q}^{\pm }\left( {\cal V}^{\mp }(\xi )\mp iv_e\frac{%
n^{^{\prime }}(\xi )}{n_0}\right) \right] .  \tag{A.2}
\end{equation}
Now, introduce another operator $\widehat{S}$, that is symmetric with
respect to operators $\widehat{Q}^{-}$ and $\widehat{Q}^{+}$, and indeed:

\begin{equation}
\widehat{S}=\widehat{Q}^{+}\widehat{Q}^{-}-\frac{\omega _L^2}{c^2}\left( 1-%
\frac{\beta _e^2}2\right) \left( \widehat{Q}^{-}+\widehat{Q}^{+}\right) +%
\frac{\omega _L^4}{c^4}\left( 1-\beta _e^2\right) .  \tag{A.3}
\end{equation}
After some transformations one can write the Eq. (A.2) with the help of
operator $\widehat{S}$

\begin{equation}
\widehat{S}{\cal V}^{\pm }(\xi )=\pm iv_e\frac{\omega _L^2}{c^2}\left[
\left( 1-\frac{\beta _e^2}2\right) \widehat{Q}^{\mp }-\frac{\beta _e^2}2%
\widehat{Q}^{\pm }-\frac{\omega _L^2}{c^2}\left( 1-\beta _e^2\right) \right] 
\frac{n^{^{\prime }}(\xi )}{n_0}.  \tag{A.4}
\end{equation}
From Eqs. (A.4) and (17) the required equation follows for variables ${\cal V%
}^{-}+{\cal V}^{+}$ and ${\cal B}^{-}+{\cal B}^{+}$:

\begin{equation}
\widehat{S}\left[ {\cal V}^{-}(\xi )+{\cal V}^{+}(\xi )\right] =-4v_e\frac{%
\omega _L^2}{c^2}\left( k_0-\beta \frac{\omega _0}c\right) \frac \partial
{\partial \xi }\frac{n^{^{\prime }}(\xi )}{n_0},  \tag{A.5}
\end{equation}
\begin{equation}
\widehat{S}\left[ {\cal B}^{-}(\xi )+{\cal B}^{+}(\xi )\right] =-4\pi
en_0\beta _e\left\{ \left( \widehat{Q}^{-}+\widehat{Q}^{+}\right) \frac
\partial {\partial \xi }+ik_0\left( \widehat{Q}^{+}-\widehat{Q}^{-}\right) -2%
\frac{\omega _L^2}{c^2}\frac \partial {\partial \xi }\right\} \frac{%
n^{^{\prime }}(\xi )}{n_0}.  \tag{A.6}
\end{equation}
Now, acting by the operator $\widehat{S}$ on both the sides of the Eq. (20)
and using the relations (A.5), (A.6), we finally arrive at the equation that
contains only the variable $n^{^{\prime }}$

\begin{equation}
\left( \frac{\partial ^4}{\partial \xi ^4}+A\frac{\partial ^2}{\partial \xi
^2}+B\right) \frac{n^{^{\prime }}(\xi )}{n_0}=-\frac{\omega _L^2}{u^2}\left( 
\frac{\partial ^2}{\partial \xi ^2}+C\right) \frac{N_b(\xi )}{n_0}, 
\tag{A.7}
\end{equation}
where

\begin{equation}
A=\frac{\omega _L^2}{u^2}\left( 1-\beta _e^2\right) +4\gamma ^4\left(
k_0-\beta \frac{\omega _0}c\right) ^2,  \tag{A.8}
\end{equation}
\begin{equation}
B=\frac{\omega _L^2}{u^2}\gamma ^2\left[ \frac{\beta _e^2\omega _L^2}{c^2}%
+4\gamma ^2\left( k_0-\beta \frac{\omega _0}c\right) ^2\right] ,  \tag{A.9}
\end{equation}
\begin{equation}
C=\gamma ^2\left[ \frac{\beta _e^2\omega _L^2}{c^2}+4\gamma ^2\left(
k_0-\beta \frac{\omega _0}c\right) ^2\right] .  \tag{A.10}
\end{equation}
One can obtain the equations for variables ${\cal E}^{\pm }$, ${\cal B}^{\pm
}$, ${\cal V}^{\pm }$, $E_z^{^{\prime }}$ in an analogous way.

Prior to the consideration of Eq. (A.7) note first of all that in the
absence of CPEM wave ($\beta _e=0)$ one has the following relation $\,$from
Eqs. (A.7)-(A.10):

\begin{equation}
\left( \frac{\partial ^4}{\partial \xi ^4}+A\frac{\partial ^2}{\partial \xi
^2}+B\right) =\left( \frac{\partial ^2}{\partial \xi ^2}+C\right) \left( 
\frac{\partial ^2}{\partial \xi ^2}+\frac{\omega _p^2}{u^2}\right) 
\tag{A.11}
\end{equation}
and Eq. (A.7) passes into the well known equation for the induced charge
obtained in Refs. [24, 25]

\begin{equation}
\left( \frac{\partial ^2}{\partial \xi ^2}+\frac{\omega _p^2}{u^2}\right)
n^{^{\prime }}(\xi )=-\frac{\omega _p^2}{u^2}N_b(\xi ).  \tag{A.12}
\end{equation}
In the limit of strong CPEM wave ($\beta _e\simeq 1$) we have from
expressions (A.7)-(A.10)

\begin{equation}
A\simeq C\simeq 4\frac{\omega _0^2}{c^2}(1+\beta )^{-2},\,\,\,B\simeq 0, 
\tag{A.13}
\end{equation}

\begin{equation}
\frac{\partial ^2}{\partial \xi ^2}\left( \frac{\partial ^2}{\partial \xi ^2}%
+q_0^2\right) n^{^{\prime }}(\xi )=0,  \tag{A.14}
\end{equation}
where $q_0=\sqrt{A}=2(\omega _0/c)(1+\beta )^{-1}$. It follows from
expression (A.14) that in this case the equation for $n^{^{\prime }}$ will
be a homogeneous differential equation. Analogous equations are obtained for
the rest of the induced variables. So, in the presence of high intensity
CPEM wave the bunch will not altogether perturb the state of homogeneous
plasma established by the external wave and all induced variables will be
zeros.

For an external wave of arbitrary intensity a non-homogeneous equation of
the fourth degree is obtained for $n^{^{\prime }}(\xi )$ (Eq. (A.7)), the
coefficients of which depend on the intensity of CPEM wave. The
characteristic equation corresponding to Eq. (A.7) establishes in general
case a dispersion law for two types of induced waves that are parametrically
coupled due to the presence of the external wave (it is easy to see that it
coincides with Eq. (33)). In case of $A^2>4B$ Eq. (A.7) describes the
oscillations of induced charge density with frequencies of $\omega _L\lambda
_{-}$ and $\omega _L\lambda _{+}$ respectively. This type of solution
corresponds to the first range of parameter values that was studied above.
In case of $A^2=4B$ the characteristic equation has multiple solutions and
respectively the solutions of Eq. (A.7) describe the waves, the amplitude of
which increases as $\xi $ (range II). At $A^2<4B$ the solutions of the
characteristic equation are complex and the waves excited by the electron
bunch exponentially decrease with the distance from the bunch (region III).

To obtain a single-valued solution of the problem one is to supplement Eq.
(A.7) with boundary conditions. As such in case of an arbitrary, smoothly
changing density profile of the bunch $N_b(\xi )$ when $\xi \rightarrow \pm
\infty $, one can use the equality of function $n^{^{\prime }}(\xi )$ and of
its first three derivatives at $\xi \rightarrow +\infty $ (or $\xi
\rightarrow -\infty $) to zero. A distinguishing feature of the
one-dimensional bunch with sharp boundaries (the profile of this kind for $%
N_b(\xi )$ we have used in calculations of induced EM fields) is the fact
that the even derivatives of the function $n^{^{\prime }}(\xi )$ are jump
functions at the boundaries of the bunch. Indeed, $n^{^{\prime }}(\xi )$ is
continuous at the boundaries of the bunch (for $\xi =0$ and $\xi =d$). As it
follows from Eq. (20), $\partial n^{^{\prime }}/\partial \xi $ is also
continuous on the boundaries of the bunch. We shall obtain by integrating
Eq. (A.7) over the variable $\xi $ in an infinitesimal range in the vicinity
of points $\xi =0$ or $\xi =d$, that $\partial ^3n^{^{\prime }}/\partial \xi
^3$ is also continuous on the boundaries of the bunch. Now consider Eq.
(A.7) in the points $\xi =\xi _0-0$ and $\xi =\xi _0+0$ (where $\xi _0=0$ or 
$\xi _0=d$) respectively. The subtraction of the obtained relations gives
the following condition:

\begin{equation}
\left( \frac{\partial ^4n^{^{\prime }}}{\partial \xi ^4}+A\frac{\partial
^2n^{^{\prime }}}{\partial \xi ^2}\right) _{\xi _0-0}^{\xi _0+0}=-\frac{%
\omega _L^2}{u^2}C\sigma n_{b,}  \tag{A.15}
\end{equation}
where $\sigma =+1$ for $\xi _0=0$ and $\sigma =-1$ for $\xi _0=d$. Besides
the condition (A.15) and the conditions of continuity of functions of $%
n^{^{\prime }}$, $\partial n^{^{\prime }}/\partial \xi $ and $\partial
^3n^{^{\prime }}/\partial \xi ^3$ at the boundaries of the bunch, there is
another boundary condition on the induced variables describing the
longitudinal waves in plasma (i.e., on $n^{^{\prime }}(\xi )$, $%
v_z^{^{\prime }}(\xi )$ and $E_z^{^{\prime }}(\xi )$). According to this
condition, owing to the equality of phase velocities of longitudinal waves
induced in the plasma to the velocity of bunch ahead of these, the functions 
$n^{^{\prime }}(\xi )$ (and, therefore, $v_z^{^{\prime }}(\xi )$) and $%
E_z^{^{\prime }}(\xi )$ turn zero. Thus, all the mentioned boundary
conditions for a bunch with sharp boundaries together with Eq. (A.7) specify
the unique solution of Eq. (A.7). The equations and boundary conditions for
transverse induced variables (${\cal E}^{\pm }$, ${\cal B}^{\pm }$, ${\cal V}%
^{\pm }$) the solutions for which coincide with expressions (22)-(25), may
be obtained in an analogous way.

\vspace{1.0in}

{\bf References}

\begin{enumerate}
\item[{[1]}]  E. Esarey, P. Sprangle, J. Krall, and A. Ting, IEEE Trans.
Plasma Sci. {\bf 24,} 252 (1996).

\item[{[2]}]  Ya. B. Fainberg, Fiz. Plazmy {\bf 23}, 275 (1997) [Plasma
Phys. Rep. {\bf 23}, 251 (1997)].

\item[{[3]}]  A. Ts. Amatuni, S. S. Elbakian, A. G. Khachatryan, and E.V.
Sekhpossian, Part. Accel. {\bf 51,} 1 (1995).

\item[{[4]}]  P. Chen, Part. Accel. {\bf 20,} 171 (1987).

\item[{[5]}]  P. Chen, J. M. Dawson, R. W. Huff, and T. Katsouleas, Phys.
Rev. Lett. {\bf 54,} 693 (1985).

\item[{[6]}]  R. D. Ruth, A. W. Chao, P. L. Morton, and P. B. Wilson, Part.
Accel. {\bf 17,} 171 (1985).

\item[{[7]}]  R. Keinigs and M. E. Jones, Phys. Fluids {\bf 30}, 252 (1987).

\item[{[8]}]  A. Ts. Amatuni, E. V. Sekhpossian, A. G. Khachatryan, and S.
S. Elbakian, Fiz. Plazmy {\bf 21,} 1000 (1995) [Sov. J. Plasma Phys. {\bf 21}%
, 945 (1995)].

\item[{[9]}]  A. G. Khachatryan, A. Ts. Amatuni, E. V. Sekhpossian, and S.
S. Elbakian, Fiz. Plazmy {\bf 22,} 638 (1996) [Sov. J. Plasma Phys. {\bf 22}%
, 576 (1996)].

\item[{[10]}]  Ya. B. Fainberg, N. Ayzatskij, V. Balakirev et al., Proc. of
XV Int. Workshop on Charged Part. Linear Accel., 1997, Alushta, p. 16 (in
Russian).

\item[{[11]}]  S. S. Elbakian, E. V. Sekhpossian, and A. G. Khachatryan,
Preprint YerPhI-1511(11)-98 (E-Print: ps/9804017).

\item[{[12]}]  A. Ts. Amatuni, E. V. Sekhpossian, and S. S. Elbakian, Fiz.
Plazmy {\bf 12,} 1145 (1986).

\item[{[13]}]  J. B. Rosenzweig, Phys. Rev. Lett. {\bf 58}, 555 (1987).

\item[{[14]}]  J. B. Rosenzweig, IEEE Trans. Plasma Sci. {\bf 15}, 186
(1987).

\item[{[15]}]  J. B. Rosenzweig, Phys. Rev. A {\bf 38}, 3634 (1988).

\item[{[16]}]  A. G. Khachatryan, Phys. Plasmas {\bf 4}, 4136 (1997).

\item[{[17]}]  J. M. Dawson, Phys. Rev. {\bf 133}, 383 (1959).

\item[{[18]}]  A. I. Akhiezer and R. V. Polovin, Zh. Eksp. Teor. Fiz. {\bf 30%
}, 915 (1956) [Sov. Phys. JETP {\bf 3}, 696 (1956)].

\item[{[19]}]  N. E. Andreev, L. M. Gorbunov, and R. R. Ramazashvili, Fiz.
Plazmy {\bf 23}, 303 (1997) [Plasma Phys. Rep. {\bf 23}, 277 (1997)].

\item[{[20]}]  A. M. Kalmykov and N. Ya. Kotsarenko, Izvestya VUZ {\bf 19},
1481 (1976) (in Russian).

\item[{[21]}]  L. Stenflo, Plasma Physics {\bf 19}, 1187 (1977).

\item[{[22]}]  C. Max and F. Perkins, Phys. Rev. Lett. {\bf 29}, 1731 (1972).

\item[{[23]}]  C. S. Liu and V. K. Tripathi, {\it Interaction of
Electromagnetic Waves with Electron Beams and Plasmas} (World Scientific
Publishing, Singapore, 1994).

\item[{[24]}]  M. E. Jones and R. Keinigs, IEEE Trans. Plasma Sci. {\bf 15},
203 (1987).

\item[{[25]}]  Hyun-Soo Kim, S. Yi, A. Amin, and K. E. Lonngren, Phys. Rev.
E {\bf 50}, 3962 (1994).
\end{enumerate}

\newpage 

\begin{center}
{\bf FIGURE\ CAPTIONS}
\end{center}

FIG. 1. The line $\Delta =\Delta (a)$, where the expression under the square
root in Eq. (34) turns zero. The region III is contained within the lines,
the region I is outside the lines, and the curve $\Delta =\Delta (a)$ is the
region II. The plot is made for $\gamma =1.5$.

FIG. 2. The line $\Delta =\Delta (a)$ for $\gamma =20$.

FIG. 3. The line $\Delta =\Delta (a)$ for $\gamma =100$.

FIG. 4. The $\xi $-dependence of the induced longitudinal electric field in
the region I of the values of parameters. The field is measured in the units
of $10^5$V/cm, $\xi $ is measured in units of $k_p^{-1}=1.7\times 10^{-3}$cm
($z_0=k_p\xi $). The curves were calculated for the following values of
parameters: $n_0=10^{17}$cm$^{-3}$, $\,n_b=10^{14}$cm$^{-3}$, $\omega
_0=3.77\times 10^{15}\sec ^{-1}$, $\gamma =50$, $k_pd=20$. The dotted line
corresponds to the absence of CPEM wave ($E_0=0$), the dashed line
corresponds to the value $E_0=1.3\times 10^{11}$V/cm, the solid line
corresponds to the value $E_0=2.5\times 10^{11}$V/cm.

FIG. 5. The $\xi $-dependence of maximum value of induced transverse
electric field $\left( E_{\max }=\sqrt{E_r^2(\xi )+E_i^2(\xi )}\right) $in
the region I. $E_{\max }$ is measured in units of $10^9$V/cm, $\xi $ in
units of $1.7\times 10^{-3}$cm. The thickness of the bunch is $10k_p^{-1}$.
The dotted line corresponds to $E_0=6.7\times 10^{10}$V/cm, the dashed line
to $E_0=8.8\times 10^{10}$V/cm, the solid line to $E_0=1.09\times 10^{11}$%
V/cm. The remaining parameters coincide with parameters given in Fig. 4.

FIG. 6. The $\xi $-dependence of field $E_r(\xi )$ (in units of $10^5$V/cm,
and $\xi $ in units of $k_p^{-1}=1.7\times 10^{-3}$cm). The parameters and
notations are the same as in Fig. 5.

FIG. 7. The $\xi $-dependence of field $E_i(\xi )$ (in units of $10^9$V/cm,
and $\xi $ in units of $k_p^{-1}=1.7\times 10^{-3}$cm). The parameters and
notations are the same as in Fig. 5.

FIG. 8. The $\xi $-dependence of the maximum value of induced magnetic field 
$\left( B_{\max }=\sqrt{B_r^2(\xi )+B_i^2(\xi )}\right) $ (in units of $10^9$%
V/cm, and in units of $k_p^{-1}=1.7\times 10^{-3}$cm). The parameters and
notations are the same as in Fig. 5.

FIG. 9. The $\xi $-dependence of magnetic field $B_r(\xi )$ (in units of $%
10^9$V/cm, and $\xi $ in units of $k_p^{-1}=1.7\times 10^{-3}$cm). The
parameters and notations are the same as in Fig. 5.

FIG. 10. The $\xi $-dependence of $B_i(\xi )$ (in units of $10^5$V/cm, and $%
\xi $ in units of $k_p^{-1}=1.7\times 10^{-3}$cm). The parameters and
notations are the same as in Fig. 5.

FIG. 11. The $\xi $-dependence of induced longitudinal electric field in the
region II (in units of $10^6$V/cm, and $\xi $ in units of $%
k_p^{-1}=1.7\times 10^{-3}$cm). The curves were obtained for values of
parameters: $n_b=10^{14}$cm$^{-3}$, $\,n_0=10^{17}$cm$^{-3}$, $\omega
_0=3.77\times 10^{15}\sec ^{-1}$, $\gamma =210$, $\,k_pd=2$. The dotted line
corresponds to the value $E_0=8.5\times 10^8$V/cm, the solid line
corresponds to the value $E_0=1.85\times 10^{10}$V/cm.

FIG. 12. The $\xi $-dependence of $E_{\max }$ in the region II (in units of $%
10^9$V/cm, and $\xi $ in units of $k_p^{-1}=1.7\times 10^{-3}$cm) for the
values of parameters: $n_b=10^{13}$cm$^{-3}$, $E_0=1.85\times 10^{10}$V/cm.
The remaining parameters are the same as in Fig. 11.

FIG. 13. The $\xi $-dependence of fields $E_r(\xi )$, $B_r(\xi )$ in the
region II (in units of $10^9$V/cm, and $\xi $ in units of $%
k_p^{-1}=1.7\times 10^{-3}$cm) for the values of parameters: $n_b=10^{13}$cm$%
^{-3}$, $k_pd=2$. The values of remaining parameters are the same as in Fig.
11. The dotted line corresponds to the value of $E_r(\xi )$, the solid line
corresponds to $B_r(\xi )$.

FIG. 14. The $\xi $-dependence of fields $E_i(\xi )$, $B_i(\xi )$ in the
region II (in units of $10^9$V/cm, and $\xi $ in units of $%
k_p^{-1}=1.7\times 10^{-3}$cm). The parameters are the same as in Fig. 13.
The dotted line corresponds to $E_i(\xi )$, the solid line corresponds to $%
B_i(\xi )$.

FIG. 15. The $\xi $-dependence of the induced longitudinal field in the
region III (in units of $10^2$V/cm, and $\xi $ in units of $%
k_p^{-1}=1.7\times 10^{-3}$cm). The parameters are: $n_b=10^{14}$cm$^{-3}$, $%
\,n_0=10^{17}$cm$^{-3}$, $\omega _0=3.77\times 10^{15}\sec ^{-1}$, $\gamma
=210$, $\,k_pd=5$. The dotted line corresponds to the value $E_0=4.09\times
10^9$V/cm, the dashed line corresponds to $E_0=1.16\times 10^{10}$V/cm, the
solid line corresponds to $E_0=1.6\times 10^{10}$V/cm.

FIG. 16. The $\xi $-dependence of $E_{\max }$ in the region III (in units of 
$10^8$V/cm, and $\xi $ in units of $k_p^{-1}=1.7\times 10^{-3}$cm) for $%
n_b=10^{13}$cm$^{-3}$. The remaining parameters and notations are the same
as given in Fig. 15.

FIG. 17. The $\xi $-dependence of the field $E_r(\xi )$ in region III (in
units of $10^8$V/cm, and $\xi $ in units of $k_p^{-1}=1.7\times 10^{-3}$cm)
for $n_b=10^{13}$cm$^{-3}$. The remaining parameters and notations are the
same as given in Fig. 15.

FIG. 18. The $\xi $-dependence of the field $E_i(\xi )$ in region III (in
units of $10^8$V/cm, and in units of $k_p^{-1}=1.7\times 10^{-3}$cm) for $%
n_b=10^{13}$cm$^{-3}$. The remaining parameters and notations are the same
as given in Fig. 15.

FIG. 19. The field $B_{\max }(\xi )$ in the region III. The values of
parameters, measurement units and notations are the same as in Fig. 16.

FIG. 20. The field $B_r(\xi )$ in the region III. The values of parameters,
measurement units and notations are the same as in Fig. 16.

FIG. 21. The field $B_i(\xi )$ in the region III. The values of parameters,
measurement units and notations are the same as in Fig. 16.

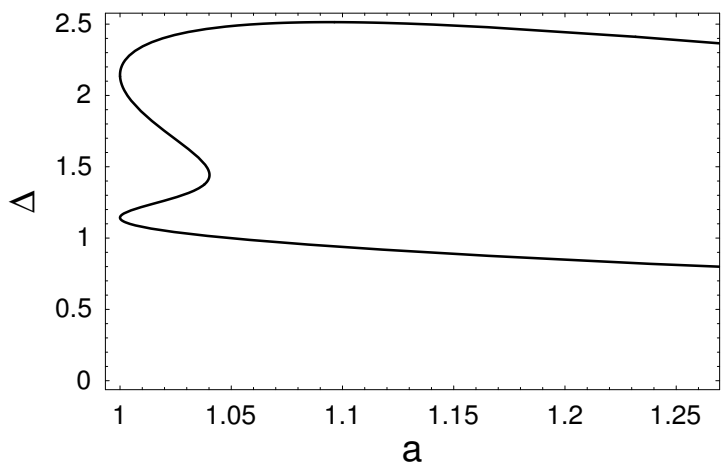
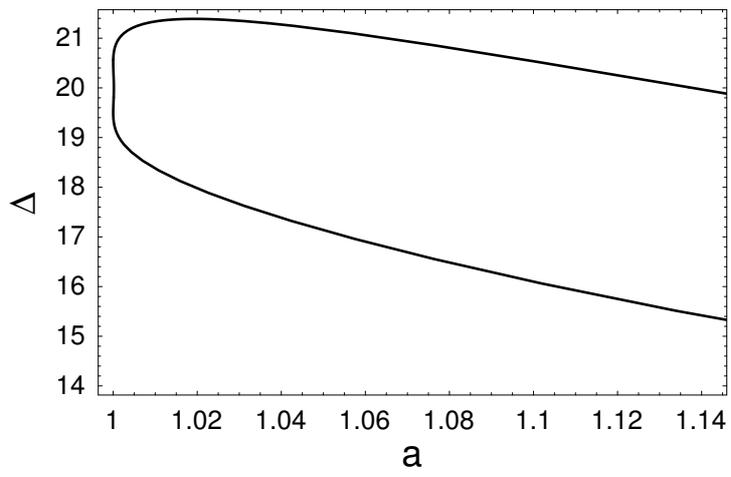
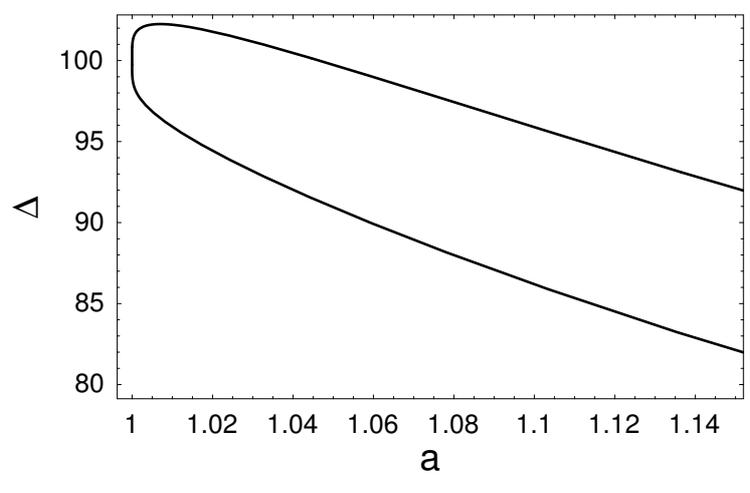
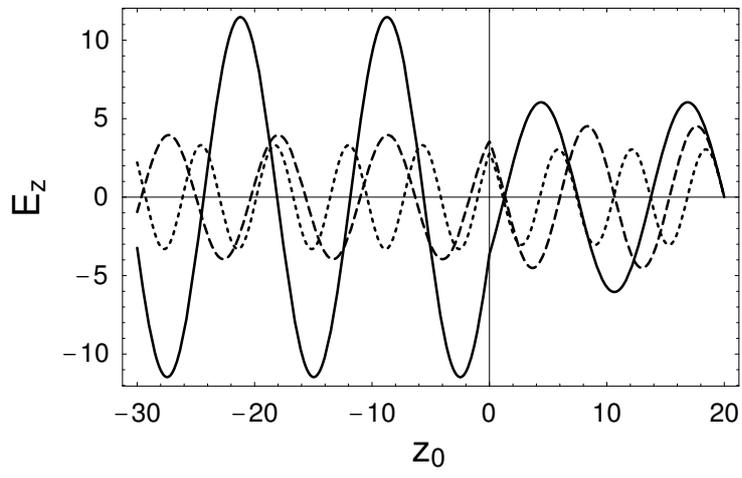
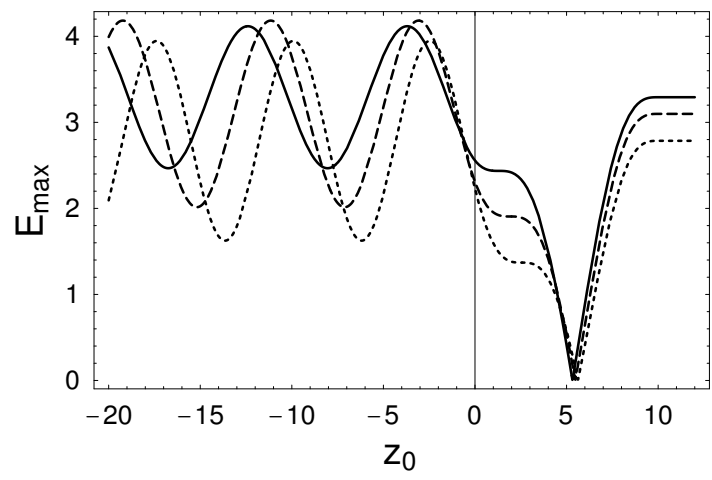
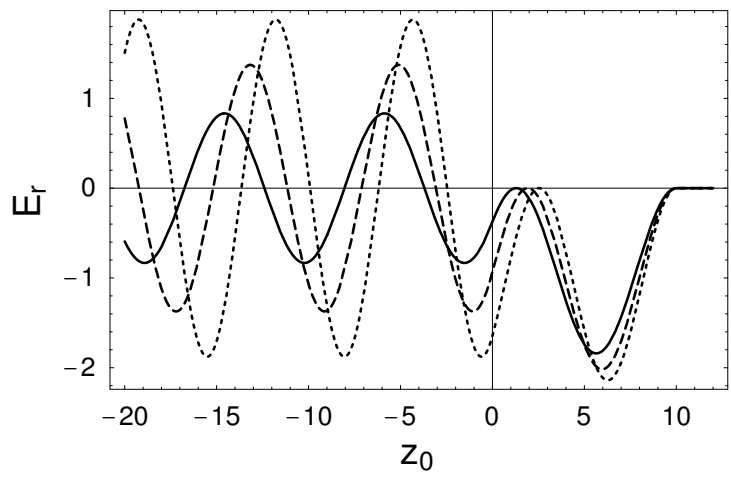
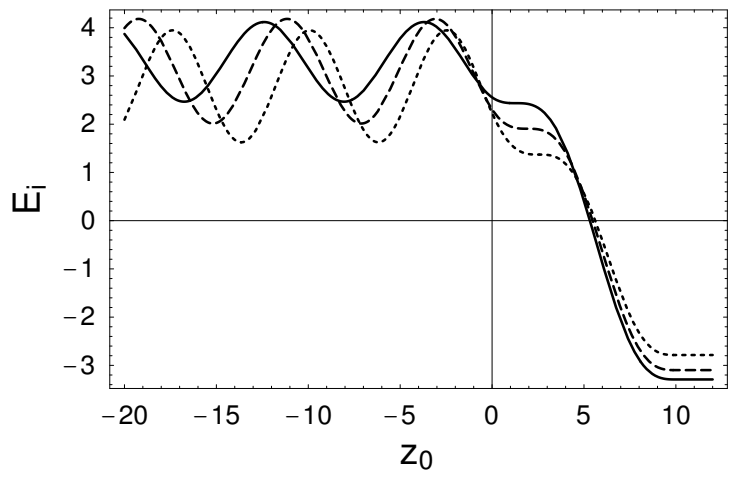
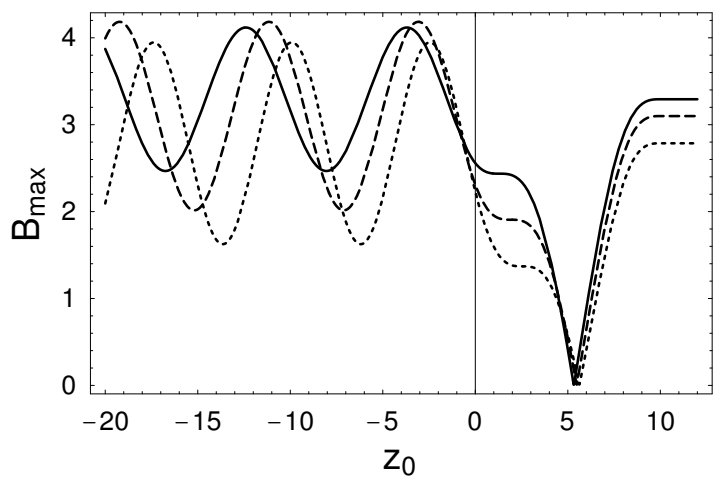
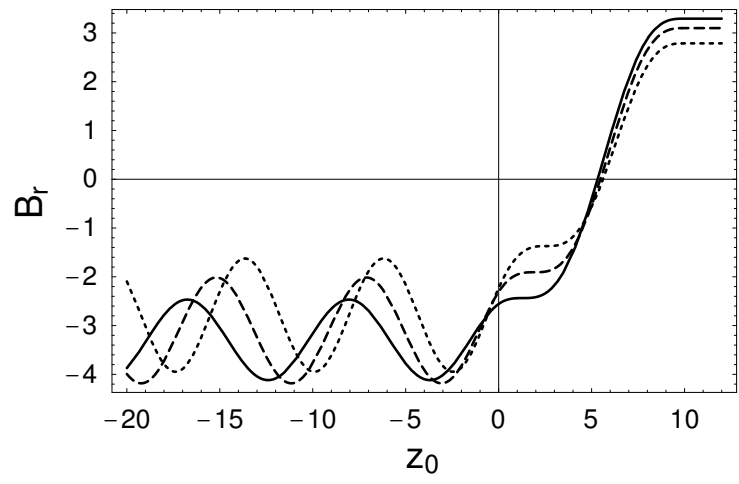
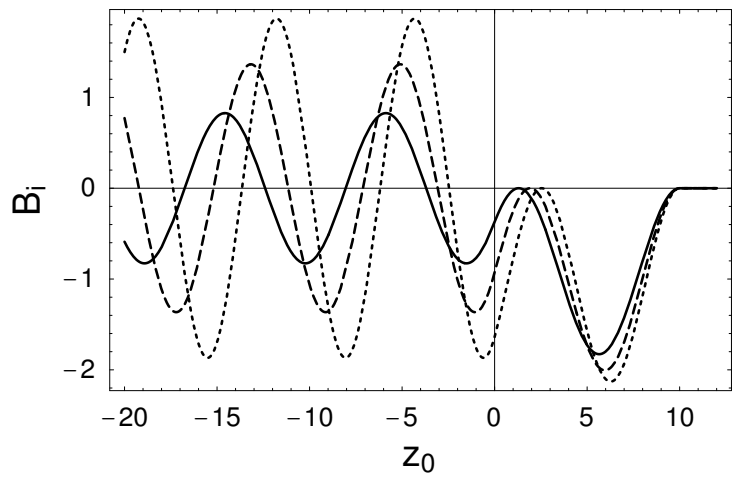
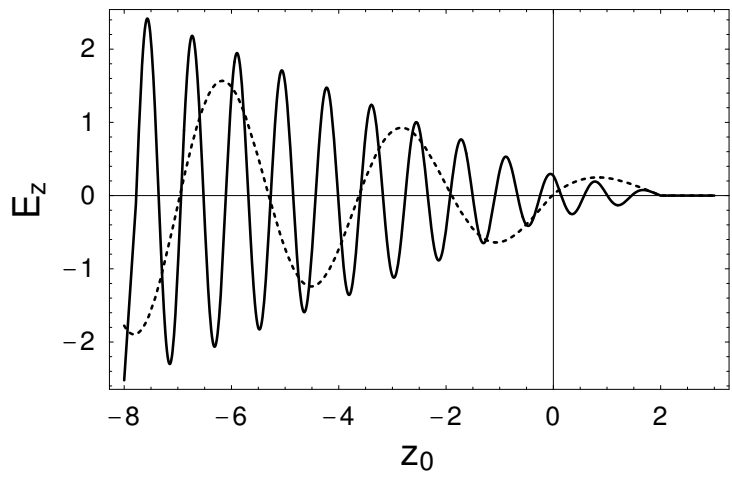
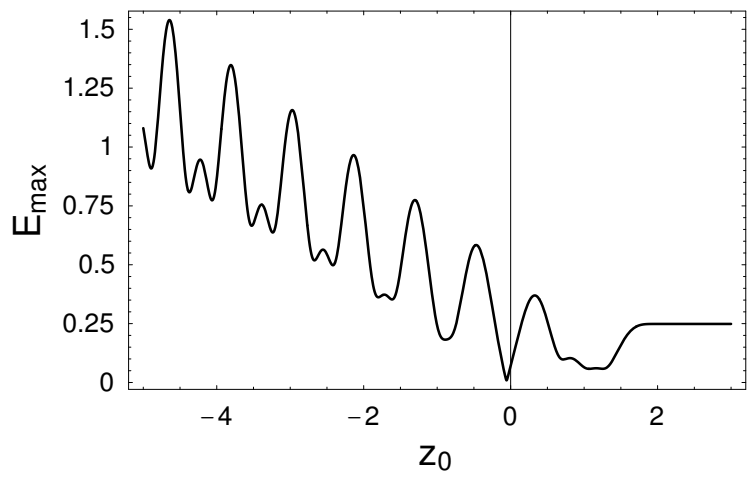
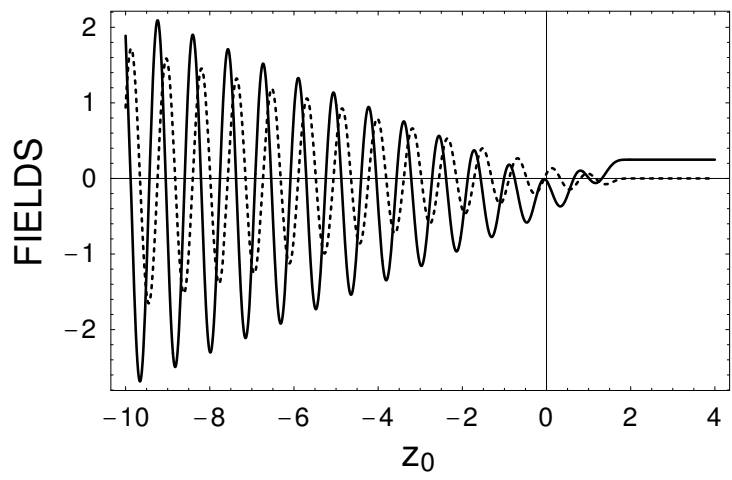
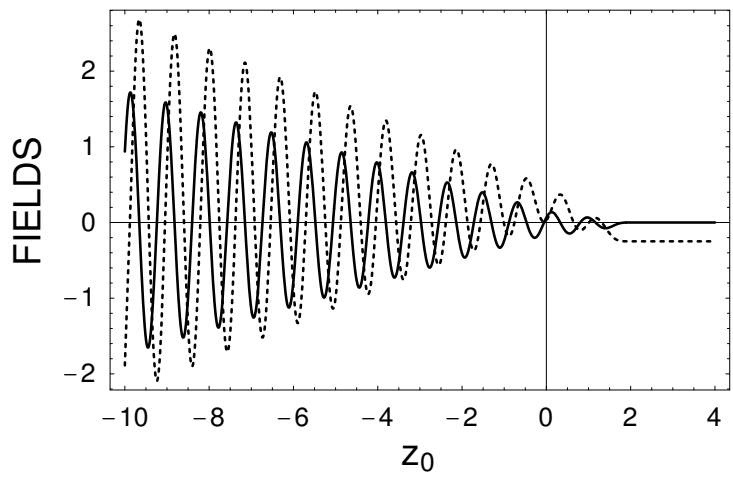
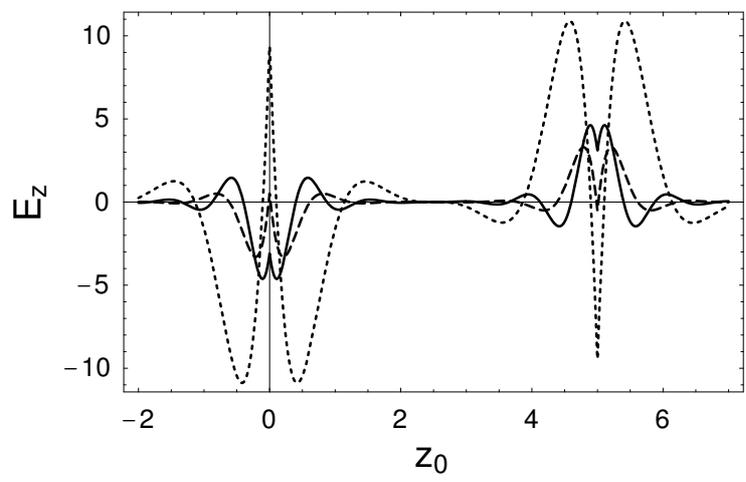
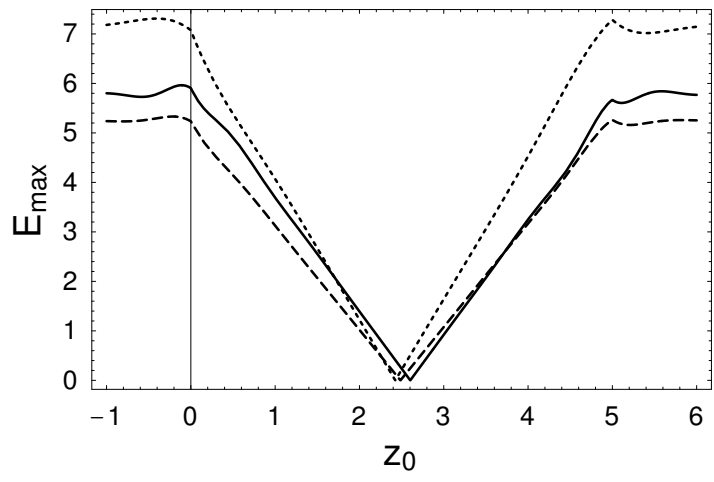
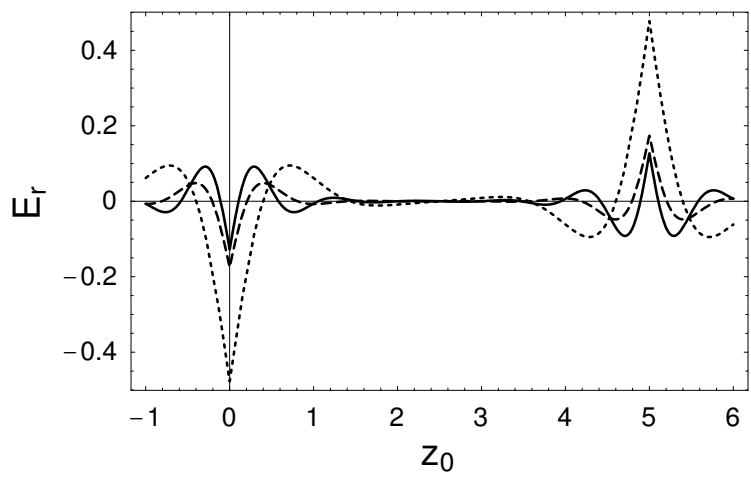
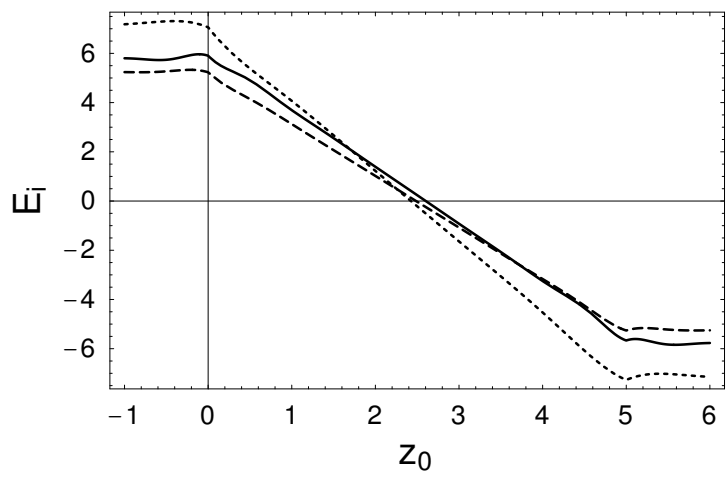
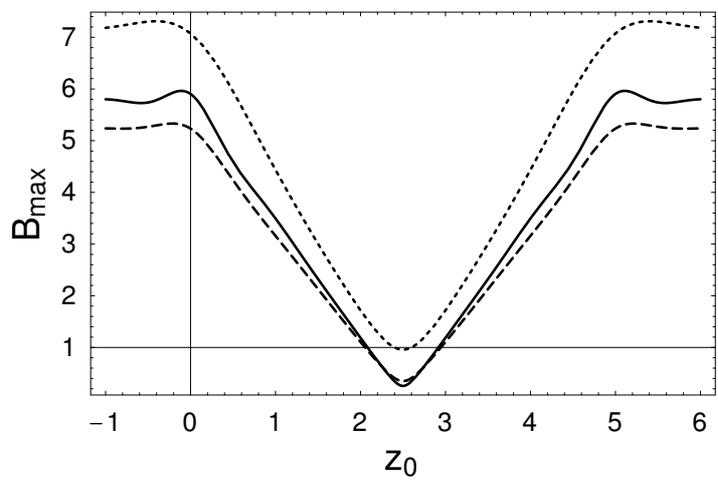
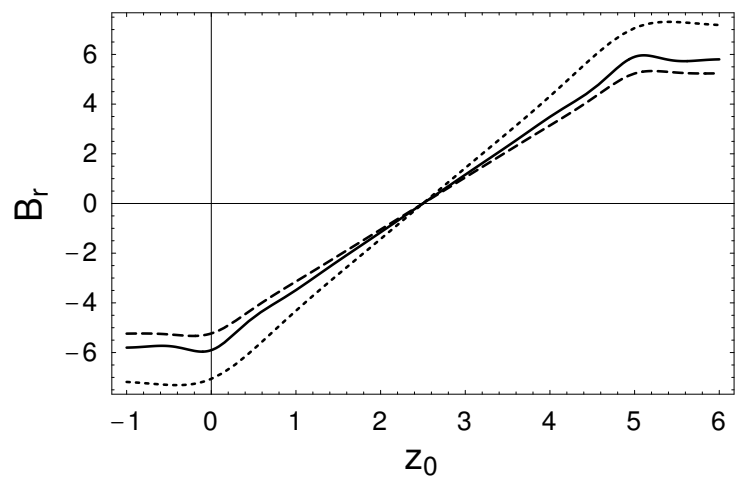
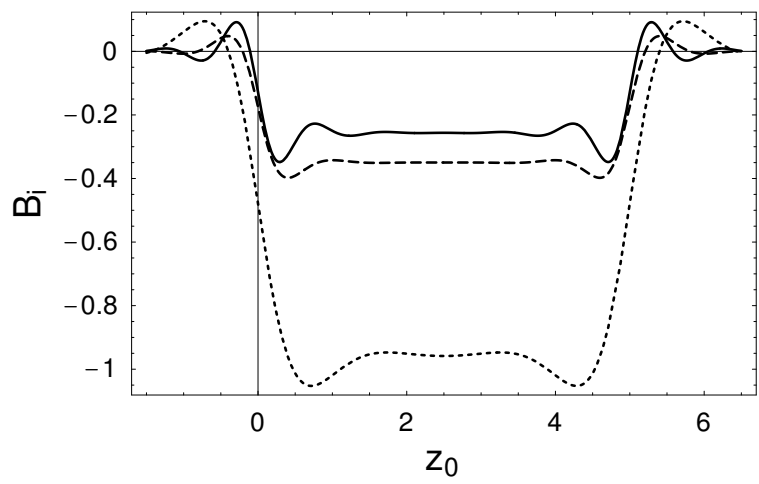
\end{document}